\documentclass[12pt,preprint]{aastex}
\RequirePackage{color}
\RequirePackage{code}
\input{astro.sty}

\usepackage[T1]{fontenc}% (2) specify encoding

\shorttitle{PSF-Matching in PS1}
\shortauthors{P.A. Price et al}

\begin{document}

\def\ale{\mathrel{\hbox{\rlap{\hbox{\lower4pt\hbox{$\sim$}}}\hbox{$<$}}}}
\def\age{\mathrel{\hbox{\rlap{\hbox{\lower4pt\hbox{$\sim$}}}\hbox{$>$}}}}

\title{Pan-STARRS PSF-Matching for Subtraction and Stacking}
% this is a crude trick to get the order of affiliations right.  These
% names are used in the affiliations below.  The user needs to (1) set
% the order and numbers to have the correct sequence in the author
% list and (2) re-order the list at the bottom (and comment-out as needed)
\def\IfA{1}
\def\Princeton{2}
\def\DUR{3}
\def\CfA{2}

\author{
P.~A. Price,\altaffilmark{\Princeton}
Eugene A. Magnier,\altaffilmark{\IfA}
}

\altaffiltext{\IfA}{Institute for Astronomy, University of Hawaii, 2680 Woodlawn Drive, Honolulu HI 96822}
\altaffiltext{\Princeton}{Department of Astrophysical Sciences, Princeton University, Princeton, NJ 08544, USA}

%% \affil{Institute for Astronomy, University of Hawaii}
%% \affil{2680 Woodlawn Drive, Honolulu, HI 96822}
%% \email{price@astro.princeton.edu, eugene@ifa.hawaii.edu}
%% 
%% \slugcomment{To be submitted to PASP} 
%% \shorttitle{Pan-STARRS PSF-matching}
%% \shortauthors{Price \& Magnier}

\begin{abstract}
We present the implementation and use of algorithms for matching
point-spread functions (PSFs) within the Pan-STARRS Image Processing
Pipeline (IPP).  PSF-matching is an essential part of the IPP for the
detection of supernovae and asteroids, but it is also used to
homogenize the PSF of inputs to stacks, resulting in improved
photometric precision compared to regular coaddition, especially in
data with a high masked fraction.  We report our experience in
constructing and operating the image subtraction pipeline, and make
recommendations about particular basis functions for contructing the
PSF-matching convolution kernel, determining a suitable kernel,
parallelisation and quality metrics.  We introduce a method for
reliably tracking the noise in an image throughout the pipeline, using
the combination of a variance map and a ``covariance pseudo-matrix''.
We demonstrate these algorithms with examples from both simulations
and actual data from the Pan-STARRS~1 telescope.
\end{abstract}

\keywords{Surveys:\PSONE, Data Analysis and Techniques}

\section{Introduction}

The past three decades have seen the increasing importance of
time-domain surveys in astronomy.  These include asteroid searches
such as the Lincoln Near-Earth Asteroid Research
\citep[LINEAR][]{2000Icar..148...21S}, the Lowell Observatory
Near-Earth Object Search \citep[LONEOS,][]{1995DPS....27.0110B}, the
Catalina Sky Survey \citep{2003DPS....35.3604L}, and ATLAS
\citep{2018PASP..130f4505T}; microlensing surveys such as MACHO
\citep{1993ASPC...43..291A} and Optical Gravitational Lens Experiment
\citep[OGLE,][]{1992AcA....42..253U}; and searches for supernovae and
other transient sources such as ASAS-SN \citep{2014ApJ...788...48S}, the
Palomar Transient Factory \citep[PTF,][]{2009PASP..121.1395L}, and the Robotic Optical
Transient Search Experiment \citep[ROTSE-I,][]{2000ApJ...542..251A}.

The Pan-STARRS Observatory \citep{chambers2017} has been a leader in
the searches for both explosive transient / supernova and potentially
hazardous asteroids.  According to the statistics maintained by David
Bishop\footnote{http://www.rochesterastronomy.org/snimages/archives.html},
since 2009, 40\% of all supernova have been discovered by
Pan-STARRS\,1.  Similarly, 24\% of all Near Earth Objects (NEOs)
discovered to date have been found by
Pan-STARRS\footnote{https://cneos.jpl.nasa.gov/stats/site\_all.html}.
Since 2014, when Pan-STARRS shifted its primary mission to the search
for NEOs, this fraction has increased to 41\%. Both of these search
programs use nightly observations to hunt for features which have
changed, either between multiple images in a single night or between
the current image and an archival reference image.

PSF-matched image differencing\footnote{We eschew the popular term
  ``difference imaging'' since we are not truly imaging differences.}
provides a useful means of identifying such variable and transient
sources, especially those superimposed on complex backgrounds such as
the Milky Way Bulge, or other galaxies.  Because images from
ground-based (and, to a lesser extent, space-based) telescopes suffer
from a variable point-spread function (PSF), mere cross-correlation of
catalogs cannot match the power of PSF-matched image differencing to
find variable sources that are faint or blended with other objects.

\citet{1998ApJ...503..325A} first proposed a fast algorithm to solve
for a convolution kernel that, when applied to an image, matches its
PSF to another image.  Though various features have been added to this
algorithm, it is still at the heart of most image subtraction codes
today.  The key to the speed of the algorithm lies in the expression
of the convolution kernel as a linear combination of basis functions,
which allows the least-squares problem to be reduced to a matrix
equation.  \citet{2000AAS..144..363A} showed how this can be expanded
to allow spatial variation of the kernel across the images.  Of
course, the basis functions used for the kernel may be completely
arbitrary.  These authors advocated using a set of Gaussians
multiplied by polynomials; \citet{2008MNRAS.386L..77B} suggests using
a set of discrete functions.  We have tried this and other
possibilities, and we will make our own suggestions for basis
functions later.

\citet{2008ApJ...677..808Y} expanded upon the original formulation of
\citeauthor{1998ApJ...503..325A} to allow for the convolution of both
images to a common PSF.  This technique provides additional
flexibility required for good subtractions when the images are
elongated in orthogonal directions, such as might be produced from
optical distortions or observing at high airmass.  It also removes the
sometimes difficult choice of which image should be convolved to match
the other (without deconvolving).

% Gal-Yam CPM: makes final PSFs larger than necessary, introducing additional noise
% Kerins et al., 2010arXiv1004.2166K: separating background and PSF matching for high surface brightness
% Albrow et al., 2009MNRAS.397.2099A: Application of Bramich to PLANET pipeline
% Quinn, Clocchiatti \& Hamuy, 2010MNRAS.403L...1Q: slow slow slow version of Alard 2000

In this paper, we outline the use of PSF-matching in the Image
Processing Pipeline (IPP) developed for the Pan-STARRS project.
First, we introduce our formulation of the problem, built upon the
above contributions (\S\ref{sec:formulation}).  Then we discuss our
implementation along with lessons learned in the development of our
PSF-matching code and its application to image subtraction
(\S\ref{sec:implementation}).  Next, we introduce a noise model which
can accurately track the noise in an image through all stages of the
pipeline (\S\ref{sec:noise}).  We introduce an application of
PSF-matching to image stacking (\S\ref{sec:stacking}) before bringing
everything together with an example SN detection (\S\ref{sec:all}).

\section{Formulation}
\label{sec:formulation}

Given two images, $I_1(x,y)$ and $I_2(x,y)$, we desire to solve for
convolution kernels, $K_1(u,v)$ and $K_2(u,v)$, such that
\begin{equation}
K_1(u,v) \otimes I_1(x,y) + K_2(u,v) \otimes I_2(x,y) + f(x,y) = 0
\end{equation}
where $f(x,y)$ is a smooth function to match the backgrounds.
Following \citet{1998ApJ...503..325A}, we will write the
background-matching function as a linear combination of basis
functions, $f_i(x,y)$, and each kernel as a linear combination of
basis functions, $g_i(x,y) k_i(u,v)$, where the inclusion of
$g_i(x,y)$ allows for spatial variation of the kernel.  In order to
enforce conservation of flux \citep[following][]{2000AAS..144..363A},
we specify that all of the kernel basis functions have zero sum,
$\sum_{u,v} k_i(u,v) = 0\ \forall i$.  This may be achieved by scaling
and subtracting one of the other kernel basis functions or simply the
central element, $\delta(u,v)$ where $\delta$ is the discretised Dirac
delta function in two dimensions, from each of the kernel basis
functions that have non-zero sum.  Then, dropping the function
variables for brevity, we seek to minimise:
\begin{equation}
\chi^2 = \sum_{x,y} \left(
b_0 I_1 +
\sum_i b_i g_i k_i \otimes I_1 +
\sum_i c_i g_i k_i \otimes I_2 +
\sum_i d_i f_i -
I_2
\right)^2 / \sigma^2 + 
\sum_i b_i^2 p_i + \sum_i c_i^2 p_i
\end{equation}
Here, the $b_i$, $c_i$ and $d_i$ are the coefficients that will match
the PSFs, and the $p_i$ are penalty functions for minimising the size
of the common PSF.  These penalty functions are necessary in the
presence of noise, since otherwise there would be little difference in
$\chi^2$ between wide and narrow kernels.  Setting $c_i \equiv 0$ and
$p_i \equiv 0$ reduces the above equation to the
\citet{2000AAS..144..363A} formalism, but with the normalisation
($b_0$) included explicitly.  In practise, the above sum will only be
over small regions (known as ``stamps''), and if we assume that the
spatial variation is not large, then we can simply use the coordinates
of the stamp centres for the $g_i(x,y)$; this allows a faster
calculation \citep{2000AAS..144..363A}.

To simplify the equation, we write
\begin{equation}
\chi^2 = \sum_{x,y} \left( \sum_i a_i A_i(x,y) - I_2(x,y) \right)^2 / \sigma(x,y)^2 + \sum_i a_i^2 P_i
\end{equation}
where we have concatenated the vectors:
\begin{equation}
\begin{array}{lcrcccccccl}
\vec{a} & = & ( & b_0 & b_i                & \ldots & c_i                & \ldots & d_i & \ldots & ) \\
\vec{A}(x,y) & = & ( & I_1 & g_i k_i \otimes I_1 & \ldots & g_i k_i \otimes I_2 & \ldots & f_i & \ldots & ) \\
\vec{P} & = & ( & 0   & p_i                & \ldots & p_i                & \ldots & 0 & \ldots & ) \\
\end{array}
\end{equation}

Minimising $\chi^2$ reduces to the matrix equation, $M \vec{a} =
\vec{v}$, where
\begin{eqnarray}
M_{ij} & = & \sum_{x,y} A_i(x,y) A_j(x,y) / \sigma(x,y)^2 + P_i \delta_{ij} \\
v_i   & = & \sum_{x,y} A_i(x,y) I_2(x,y) / \sigma(x,y)^2
\end{eqnarray}
Here $\delta_{ij}$ is the Kronecker delta.  Once this equation has
been solved using regular matrix methods, the convolution kernels and
background matching function may be easily calculated and applied.
Note that, with the above formulation, the convolution kernel to be
applied to $I_2$ so that the two convolved images may be subtracted is
$K_2(u,v) = \delta(u,v) - \sum_i c_i g_i k_i$.

Up until now, we have been restricting ourselves to a very general
formulation.  We now point out some specific choices to be made.
Adopting $x^\ell y^m$ (i.e., ordinary polynomial) or a Chebyshev or
other special polynomial for the $f_i(x,y)$ and $g_i(x,y)$ is simple
and convenient.  The kernel basis function sets of
\citet{1998ApJ...503..325A} and \citet{2000AAS..144..363A} are
\begin{equation}
g_i(x,y) k_i'(u,v) = \psi_i x^\ell y^m u^p v^q \exp((u^2+v^2)/2s_i^2)
\end{equation}
for integer values of $\ell,m,p,q$, and where the $\psi_i$ is a
normalisation constant, $s_i$ is a Gaussian width, and the prime on
$k_i$ indicates that it has not been normalised to zero sum (contrary
to the above requirements, but shown like this for simplicity).
\citet{2008MNRAS.386L..77B}'s choice of kernel basis functions is
\begin{equation}
g_i(x,y) k_i'(u,v) = \delta(u-\ell,v-m)
\end{equation}
(where spatial variation is implemented by applying this separately to
discrete areas).  Other basis functions are possible, and we will
elaborate on these and some others we have tried in the next section.

Following \citet{2008ApJ...677..808Y}, we use as the penalty function
\begin{equation}
P_i = \Phi \sum_{u,v} (u^2 + v^2)^2 k_i(u,v)^2
\end{equation}
where $\Phi$ is chosen so that the $P_i$ have a contribution to the
$A_{ii}$ that is neither negligible nor dominant (so that the equation
is permitted to feel the force of the penalty, without being
overwhelmed).

\section{Implementation}
\label{sec:implementation}

Our implementation of PSF-matching is in the \code{psModules} library,
which is wrapped by the \code{ppSub}\footnote{The \code{pp} in the
  name stands for a once-proposed name for the Pan-STARRS IPP,
  ``Pan-Pipes'', and not the name of the first author.} program.  Both
are contained within the IPP software tree.

\subsection{Kernel Basis Functions}

The fact that the equations for PSF-matching can be written in a
general manner  (i.e., the choice of a different kernel basis set is
not a different algorithm, but only a different functional choice)
indicates that it is not difficult to code a system that will allow
multiple basis functions.  We have implemented a number of kernel
basis function sets in \code{psModules} in an attempt to improve the
quality of image subtractions.  In the course of this, we have
determined the following qualities to be desirable:
\begin{enumerate}
\item Generally smooth.  If the basis functions have significant
  high-frequency power, the final convolution kernel(s) may introduce
  excess noise into the image.  This is especially important in the
  outer part of the kernel, since this region is not as well
  constrained as the inner parts by the stamps.
\item Trending to zero.  A basis function that is not trending to zero
  at the edges introduces a discontinuity which will result in
  additional noise and/or artifacts in the image (e.g., a ring around
  a point source).
\item Span the space.  If a linear combination of the basis functions
  cannot reproduce the actual kernel(s) required for matching the two
  PSFs, then the subtraction will produce systematic residuals.
\item Small number of parameters.  A large number of parameters
  increases the cost of computation and adds to the possibility of
  introducing extra noise.
\end{enumerate}

The \code{POIS}\footnote{Pan-STARRS Optimal Image Subtraction} kernel
basis functions in \code{psModules} are equivalent to the
$\delta$-functions advocated by \citet{2008MNRAS.386L..77B}.  Despite
their tremendous flexibility (quality \#3), these basis functions fail
the other three desirable qualities.  In particular, they are at the
extreme opposite end of the spectrum of smoothness, and the number of
parameters required for a kernel to match images with rather different
PSF widths becomes prohibitive.  In an attempt to redeem the notion of
discrete basis functions that completely span the space without the
need to specify additional parameters, we attempted binning the
convolution kernel, so that the basis functions were composed of
multiple clustered $\delta$-functions, with the number within a
cluster growing with distance from the centre.  We found that this
reduced the flexibility, while failing to address the issue of
smoothness.  Binning radially (using rings, as opposed to blocks)
suffered the same problems.

% \tbd{Gene: add discussion of Hermitians and deconvolved Hermitians.}

We have therefore been driven back to the classical \code{ISIS} kernel
basis functions introduced by \citet{1998ApJ...503..325A}.  These
satisfy all of the above qualities except for spanning the space.
However, we have found that, with a suitable choice of Gaussian
widths\footnote{We prefer to parametrise these by a Gaussian
  full-width at half maximum (FWHM) instead of the traditional
  standard deviation, since astronomers tend to be more familiar with
  FWHM.} and by scaling these widths appropriately, we find that the
\code{ISIS} kernels are quite flexible.  In particular, a Gaussian of
narrow width (e.g., FWHM $\sim 2$~pixels) is very effective at
producing the required structure in the centre of the kernel,
especially when combined with a moderately high polynomial order,
while wider Gaussians do a good job of modelling the `skirt', before
trending to zero.  Our current recommended prescription for
\code{ISIS} kernels are three Gaussians, with FWHM of 2.9, 4.8 and 8.2
pixels and (ordinary) polynomial orders of 4, 2 and 2 respectively.
Following advice from A.~Rest (priv.\ comm.), we linearly scale these
by the maximum FWHM of the inputs (attempts to scale by the square
root of the difference of the squares, as might be naively expected,
were not as effective).  Our prescription is appropriate for a maximum
FWHM of 9 pixels, and we set a minimum scale of 0.7 (to avoid too
small a Gaussian width, which can lead to high frequency noise) and a
maximum scale of 1.2 (to avoid too large a kernel, which increases
computation cost and makes it more difficult to find stamps in images
with a large masked fraction).

\subsection{Normalisation}

We have found that the most important parameter to measure accurately
is the normalisation between the two images, $b_0$ in the above
formulation.  Because the other kernel basis functions have zero sum,
an error in the normalisation cannot be compensated by tweaking the
other coefficients.  Nevertheless, through the least-squares solution,
the coefficients of what would normally be even functions (without the
zero sum) are adjusted to minimise the $\chi^2$ in an attempt to
compensate, resulting in strong ring-like residuals.

In order to accurately measure the normalisation without the
interference of the convolution kernels, we solve for $b_0$ and a
constant background differential term ($d_0$) separately from the
other parameters.  We may do this because the normalisation and
background differential are the only terms that add flux to the image;
the other terms just move the flux around.  Taking a lesson from
PSF-fitting photometry, we specifically do not weight by the variance
in the matrix accumulations, i.e., we set $\sigma(x,y) \equiv 1$.
This is because the smaller Poisson noise of brighter pixels would
bias the normalisation too high.

% \tbd{Window function?  We don't use it any more, but should we mention it?}
%% We attempted accumulating the matrix elements for these two terms
%% using an empirical window function constructed from a median of all
%% the stamps, with the outer 3\% of the total contribution removed so
%% that pixels with low signal-to-noise do not bias the measurement; this
%% is analagous to choosing a suitable aperture size for aperture
%% photometry.  However, this was not effective.

\subsection{Stamps}

Since we restrict the analysis of the kernel required for PSF matching
to the small ``stamps'' centered on bright stars, the choice of stamps
is key to successful PSF-matching.  The convolution kernel is only as
good as the stamps used to construct it.  We use a merged list of
sources from photometry of the two input images as the basis of our
stamps list.  Sources with a flag indicating that it is anything other
than a pristine astrophysical source are excluded.  At the present
time, we make no effort to select sources of a particular color or
range of colors.

We exclude sources with any masked pixels that would affect the
calculation of the convolution kernel.  For a camera such as
Pan-STARRS~1's Giga-Pixel Camera~1 (GPC1) with a high ($\sim 20\%$)
masked fraction and where these masked pixels are widely distributed,
this can make it difficult to find sufficient high-quality stamps.  To
minimise this, we try to keep the size of the convolution kernel
small.

We divide the image into subregions of size typically about
1~arcmin$^2$.  The source with the brightest peak pixel over a defined
threshold (e.g., $5\sigma$ above background) within each subregion (if
any) is nominated as a stamp center.  The use of these subregions
ensures the stamps are distributed over the entire image.

We do not explicitly discriminate against galaxies as stamps (they can
provide information on the convolution kernel, though not as
effectively as point sources), except that stars will typically be
chosen first within a subregion because of their higher surface
brightness.

Discrepant stamps are removed (and replaced, if possible) using
multiple solution iterations.

\subsection{Masking}

Masked pixels (e.g., due to saturation, artifacts, etc.) do not
contribute to the convolution.  To minimise the effect these pixels
have on the convolved image(s), we adopt a distinction between ``bad''
pixels and ``poor'' pixels first suggested by A.~Becker and
implemented in the \code{HOTPANTS}
code\footnote{\url{http://www.astro.washington.edu/users/becker/hotpants.html}}.
Pixels in a convolved image that are seriously affected by the
presence of a masked pixel in the original image are flagged as
\code{CONV.BAD}, while pixels that may only be slightly affected are
flagged as \code{CONV.POOR}.  The (box) radius of the distinction
between the two is chosen to be the radius at which the sum of the
square of the kernel drops below a chosen fraction (currently the
default is 0.2) of the total.

\subsection{Optimisations and parallelisation}

Parallelisation is becoming increasingly important as practical limits
to shrinking sizes and increasing clock rates set in, and the CPU
industry turns to multiple core architectures for increasing
computational power.  Fortunately, the two major stages of the
PSF-matching process may be easily parallelised.

The first major stage of the PSF-matching process is the calculation
of the equation.  This may be done on each of the stamps in
independent threads.  We calculate a least-squares matrix
independently for each stamp, since these need not be recalculated for
each rejection iteration.  Summing these matrices and producing a
solution is fairly fast and need not be threaded.

The second major stage is the application of the solution to the
images.  Having assumed in the construction of the least-squares
equation that spatial variations of the kernels are small on the scale
of a stamp, we are quite entitled to continue that assumption by
dividing the image into regions about the size of a stamp and applying
a single kernel to each in independent threads.  We use Fast Fourier
Transforms to perform the convolutions, as this is faster than a
direct convolution for kernels larger than about 13 pixels (full
width, square), as ours tend to be.

\subsection{Quantifying quality}

We calculate several quantities for the purposes of evaluating the
quality of the PSF-matching process.  These are stored in our
processing database, allowing a quick comparison of these quantities
for the multitude of images processed, and identification of poor
subtractions.

For each stamp, we calculate the residual from applying the solution.
In this case, we \textit{do} divide through by the variance, softened
by adding a scaled version of the flux, so that the brightest pixels
do not exert undue influence on the fit.  We use the mean of the
squared residuals (effectively a $\chi^2$) as a measure of the quality
of the stamp's subtraction.  The mean and standard deviation of these
residuals over the collection the stamps serves as a measure of the
quality of the subtraction over the entire image.  The number of
stamps used in the calculation is also recorded.

Deconvolution (narrowing a PSF instead of broadening it) is often the
enemy of the PSF-matching process --- when the solution for the kernel
involves deconvolution, the result is usually noisy or inaccurate.  We
quantify the deconvolution by measuring, as a function of radius, the
sum of kernel values enclosed within that radius relative to the total
kernel sum.  When the maximum deconvolution fraction is significantly
in excess of unity, then the result is typically unacceptable.  We
also calculate and record the first and second moments of the kernel
(realised at the center of the image, if spatially variable).

% \tbd{Add fSigRes, fMaxRes, fMinRes statistics?}

\subsection{Examples}

\begin{figure}
\plottwo{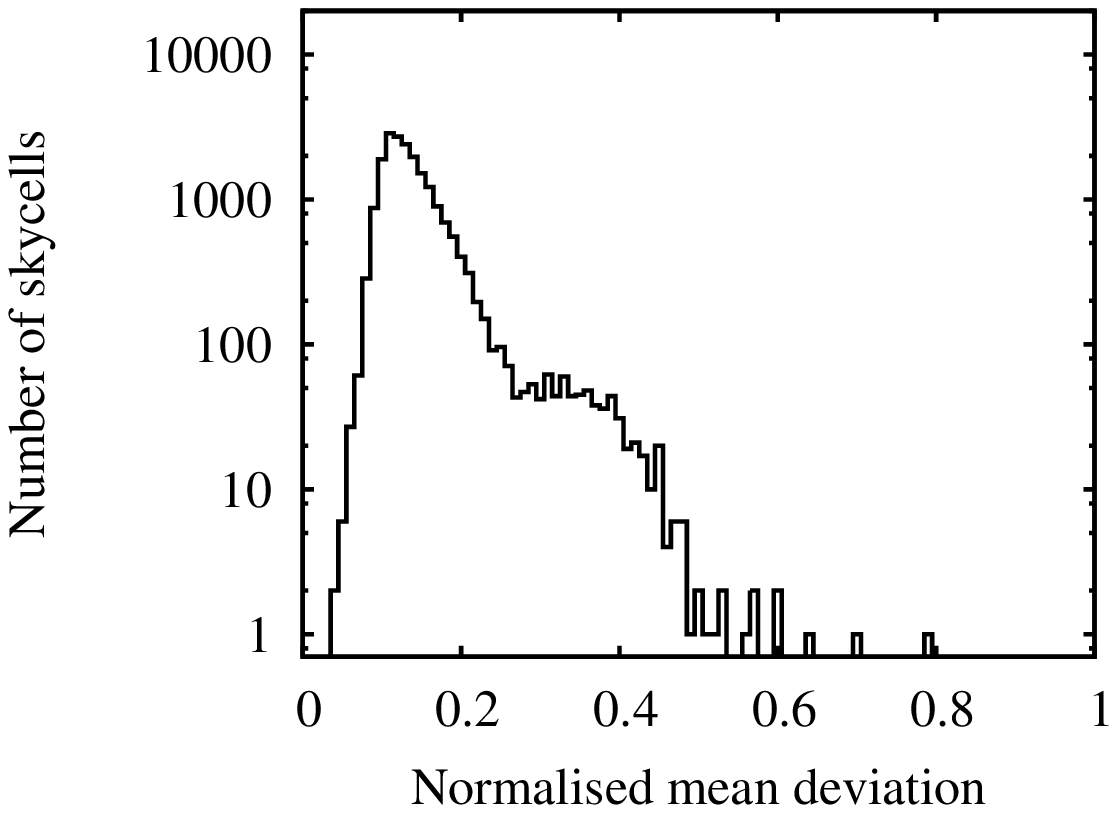}{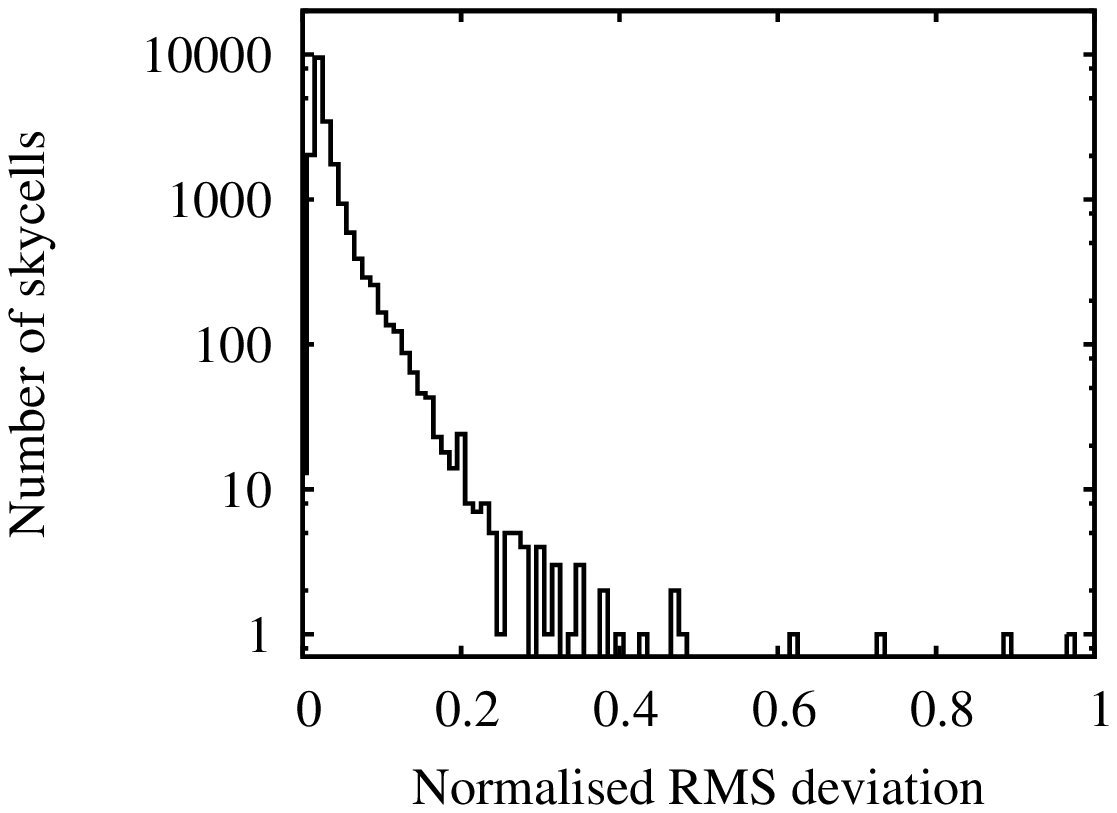}
\plottwo{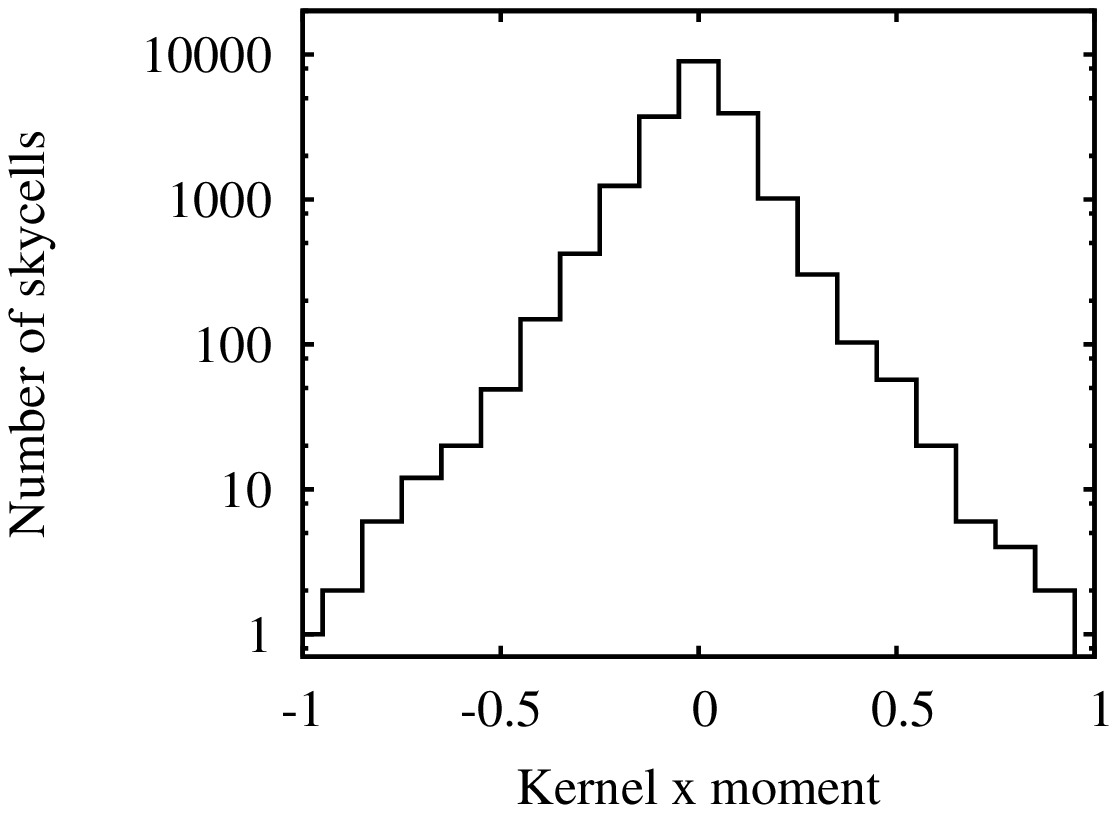}{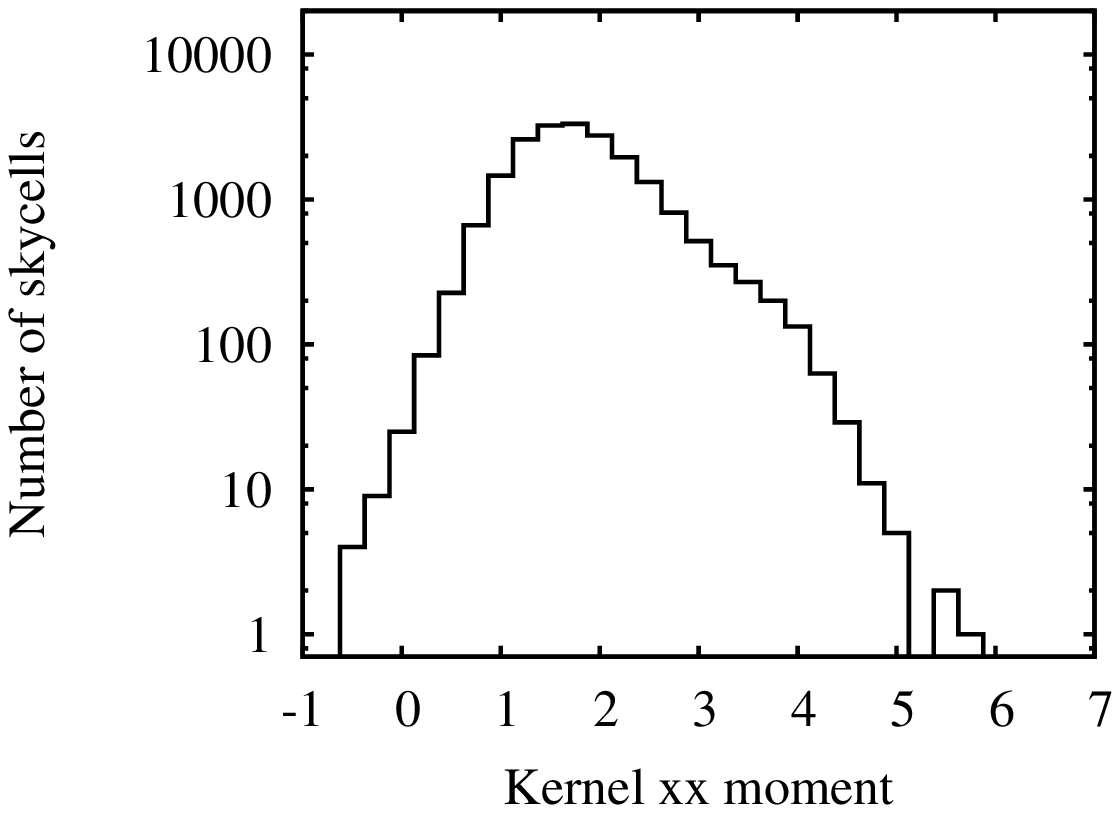}
\plottwo{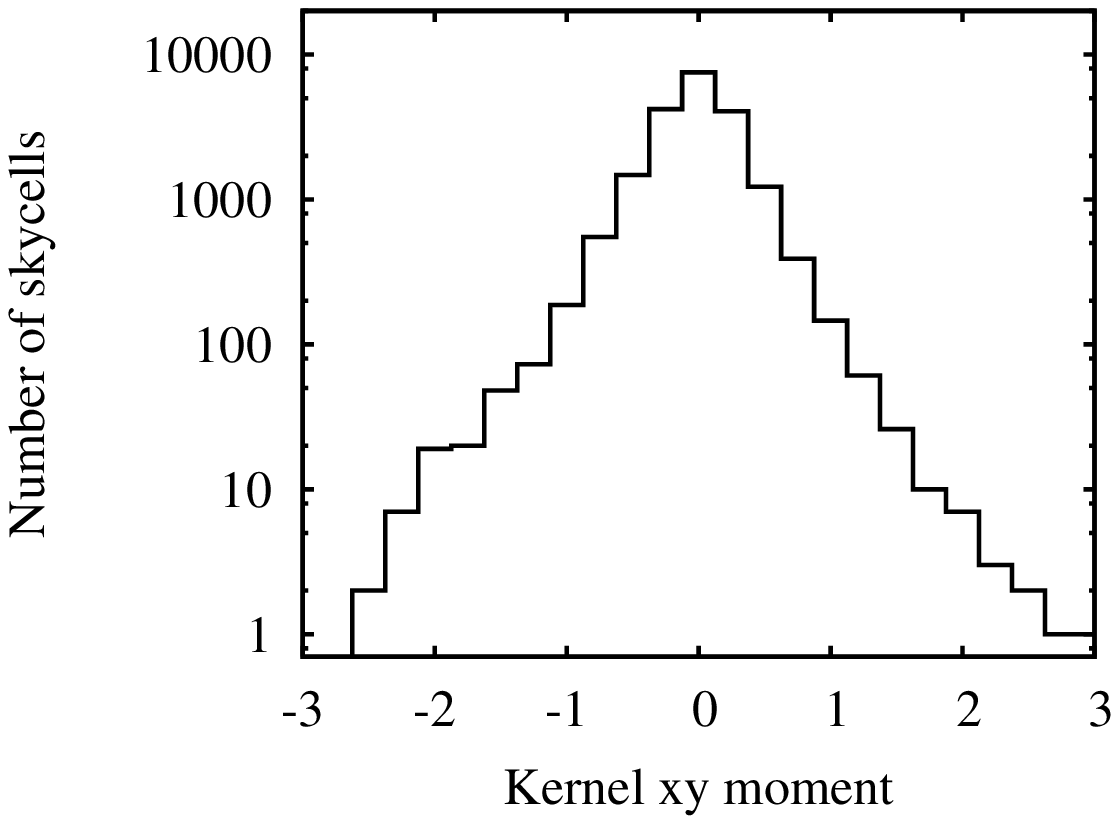}{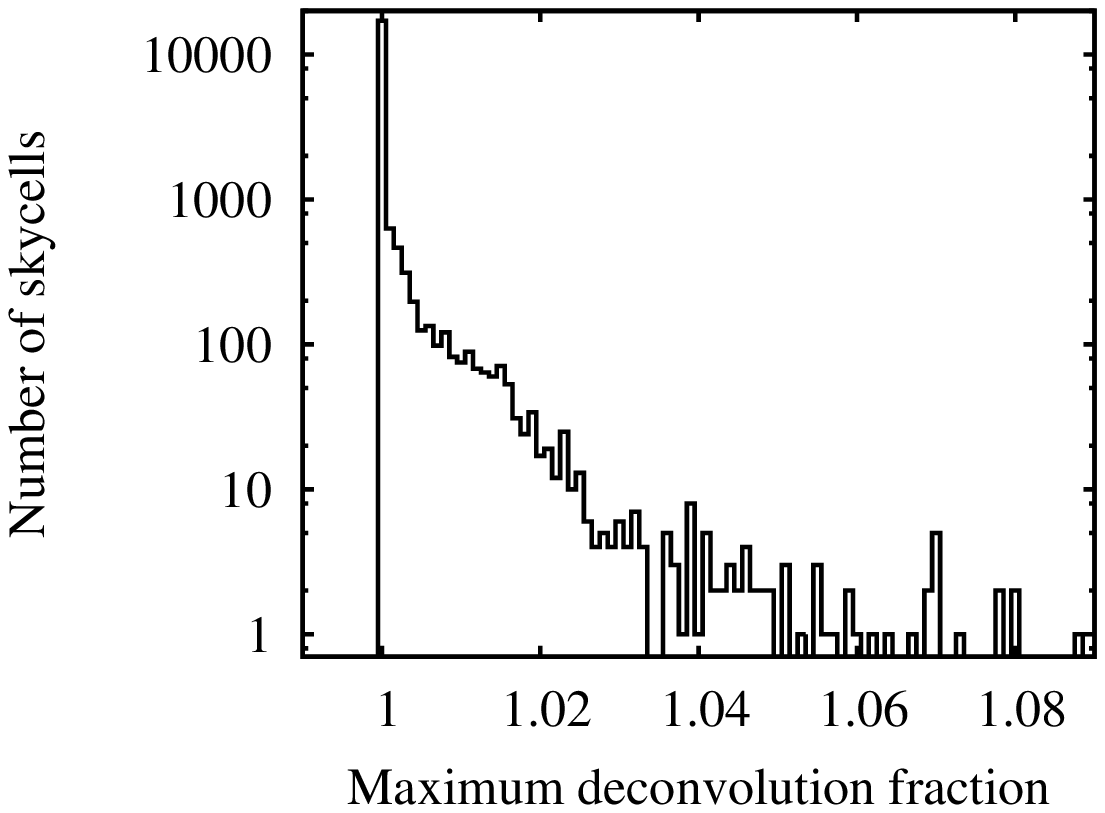}
\caption{Distributions of quality metrics from 20,053 subtractions.
  The kernel y and yy moments are similar to the x and xx moments,
  respectively, so we do not show them.}
\label{fig:diff:histograms}
\end{figure}

On the night of 2010 June 27 (UT), Pan-STARRS~1 (PS1) took
550~exposures (534 as part of the ``3$\pi$ Survey'' and 16 as part of
the ``Medium-Deep Survey'') which were automatically processed by the
IPP.  As part of this processing, 20,240 subtractions were made.  Of
these, 20,053 completed cleanly, while 187 were flagged as having bad
quality data (2 due to insufficient unmasked area, and 185 due to
failure to measure the PSF on the convolved image).  In
Figure~\ref{fig:diff:histograms} we show the distributions of various
quality metrics.

Based on these distributions and inspection of the subtractions, we
recommend using the following cuts to flag bad subtractions:
\begin{itemize}
\item Number of stamps used $<$ 3
\item Normalised mean deviation $>$ 0.5
\item Normalised r.m.s.\ deviation $>$ 0.3
\item First moments (x, y) of kernel $>$ 0.5
\item Second moments (xx, yy) of kernel $<$ 0
\item Maximum deconvolution fraction $>$ 1.03
\end{itemize}
Applying these cuts flag 132 of our subtractions (or 0.66\%) as bad.

\section{Noise Model}
\label{sec:noise}

The accurate detection and measurement of astrophysical sources on
images depends on a reliable noise model.  The usual practise in the
analysis of optical images is to assume that the noise in a processed
image is Poisson ($\sigma^2 = N + r^2$, where $\sigma$ is the noise in
electrons, $N$ is the number of electrons recorded and $r$ is the
detector read noise in electrons; the electronic gain is required to
convert between data units and electrons) or to measure the noise
directly from the image.  Both of these practises ignore the reduction
history of the image being analysed, erring in the presence of
correlated noise produced by resampling and convolving, or even when
there is non-uniform vignetting or dark current.  This has led to some
progressive software packages producing \citep[e.g.,
  SWarp:][]{2002ASPC..281..228B} and using \citep[e.g.,
  SExtractor:][]{1996AAS..117..393B} weight maps to characterise the
noise over the image.  Because of simplicity and lower calculation
cost relative to weights or standard deviations, we prefer to
calculate and propagate the variance as the ``weight map''.

\subsection{Covariance}

In the presence of multiple convolutions\footnote{An interpolation is
  effectively the same as a convolution for these purposes.}, merely
tracking the variance is insufficient to characterise the noise
properties of an image.  A convolution moves noise from the diagonal
terms (``variance'') of the covariance matrix into the off-diagonal
terms (``covariance''), so that subsequent convolutions, even if they
attempt to account for the variance, will mis-estimate the noise in
the twice-convolved image because the noise pushed into the covariance
by the first convolution has not been accounted for in the second
convolution.

Multiple convolutions are impossible to avoid for any pipeline that
wishes to do anything more than photometry of an image directly from
the detector.  If sources are to be identified on a subtracted image,
there will usually be at least one resampling/interpolation to align
the images\footnote{One might only align images using integer offsets
  to avoid resampling, but this dread of convolutions is unnecessary
  since at least two convolutions are inevitable, and their effects
  may be characterised, as we will show.}, at least one convolution to
match the PSFs and then another convolution by a matched filter to
identify sources --- three convolutions.  It is therefore important to
treat the covariance.

However, it is not feasible to calculate or store a full covariance
matrix.  For example, a $2048\times 2048$ image (about $4\times
10^6$~pixels; small by modern standards) would require in excess of
$10^{13}$~elements.  Even if we recognise that the covariance matrix
is quite sparse, the size is still prohibitive (a $6\times 6$~kernel,
typical for Lanczos interpolation, applied to our hypothetical image
would require in excess of $10^8$~elements).

Instead, we track the average covariance for an image using a
``covariance pseudo-matrix'' (of order a few~$\times 10^2$ elements),
combined with a variance map (same dimension as the flux image).  The
covariance pseudo-matrix tracks the average covariance between a
single pixel and its neighbours (i.e., the off-diagonal terms of the
full covariance matrix relative to the diagonal terms), while the
covariance map tracks the variation of the variance across the image
(i.e., the diagonal terms of the full covariance matrix).  The
covariance pseudo-matrix is assumed to be identical for all pixels;
this isn't strictly true because convolution kernels may vary over the
image, yet it's a reasonable and useful approximation that allows us
to accurately model the noise in the image through all stages of the
pipeline.

\subsection{Prescription}

When we apply a convolution kernel to an image, we are making a linear
mapping from $m$ variables, $x_i$ (the flux in the unconvolved image),
to $n$ variables, $y_j$ (the flux in the convolved image):
\begin{equation}
y_j = \sum_i A_{ij} x_i
\end{equation}
If the covariance matrix for $\vec{x}$ is $M^x$, then the covariance
matrix for $\vec{y}$
is\footnote{\url{http://en.wikipedia.org/wiki/Propagation\_of\_uncertainty}}:
\begin{equation}
M^y_{ij} = \sum_k \sum_\ell A_{ik} M^x_{k\ell} A_{\ell j}
\end{equation}

A covariance pseudo-matrix may be thought of like a kernel (i.e., an
image extending to both positive and negative offsets in both
dimensions, where the central pixel corresponds to zero offset on the
image to which it is applied), where the $0,0$ element is the
reference for all others.  To calculate an element of the covariance
pseudo-matrix of the convolved image, we sum the product of all
possible combinations of stepping from the element of interest to the
central (reference) element via a convolution kernel ($A_{ik}$), the
input covariance pseudo-matrix ($M^x_{k\ell}$) and another instance of
the convolution kernel ($A_{\ell j}$).  The output covariance
pseudo-matrix therefore has dimensions of twice the size of the
kernel, plus the size of the input covariance pseudo-matrix.

The calculation of the covariance pseudo-matrix, requiring three
nested iterations, can become expensive.  To prevent this, the input
pseudo-matrix and/or kernel may be truncated by discarding the outer
parts which are not strongly significant (e.g., the outer 1\% of the
sum of absolute values).  The calculation is also easily parallelised,
since each element of the pseudo-matrix may be calculated
independently of the others.

When convolving, the image and variance are calculated in the usual
manner:
\begin{eqnarray}
I'(x,y) & = & \sum_{u,v} k(u,v) I(x-u,y-v) \\
V'(x,y) & = & \sum_{u,v} k(u,v)^2 V(x-u,y-v)
\end{eqnarray}
where $I(x,y)$ is the (flux) image, $V(x,y)$ is the variance map,
primes indicate the result of the convolution, and $k(u,v)$ is the
convolution kernel.  Due to the separation of the variance and
covariance, there is an additional factor of $K = [\sum_{u,v}
  k(u,v)^2]^{-1}$, which may be absorbed in either the variance or
covariance pseudo-matrix (since it is multiplicative).

Then, with the covariance pseudo-matrix calculated, the
variance for each pixel on the (flux) image is the variance map
multiplied by the reference (central) value of the covariance
pseudo-matrix (the ``covariance factor'').  For convenience, we
normally transfer the covariance factor from the pseudo-matrix into
the variance map, so that analyses that do not include convolution
need not bother with the covariance.

% \tbd{Gene: Photometry, i.e., statistics on extended regions?}

\subsection{Use in practise}

When using a spatially variable kernel (e.g., interpolation by
anything more than a constant shift; or PSF-matching as outlined
above), it is impossible for this simple model (variance map with a
single covariance pseudo-matrix) to provide a perfect description of
the noise properties of an image, and approximations must be made.
Our usual practise is to calculate a sample of covariance
pseudo-matrices over the image and take the average as representative
of the entire image.  Absorbing the $K$ into the variance map washes
out the structure caused by the spatially variable kernel.  Absorbing
the $K$ into the covariance pseudo-matrix washes out changes in the
covariance as a function of position, but this is necessarily the case
anyway (there is only one covariance pseudo-matrix) and covariance
errors are a more subtle effect than variance errors, and so this is
our preference.

When applying a spatially-variable PSF-matching kernel, we apply a
constant kernel to small patches.  For each patch we calculate the
covariance factor (not the entire covariance pseudo-matrix, which
takes much longer) from the kernel and apply it to the variance of
that patch.  We then remove the covariance factor from the average
covariance pseudo-matrix (calculated from a much smaller array of
kernels, because of the expense of calculation) since it has already
been applied.

Binning an image may be thought of as convolution where the kernel has
a constant value, followed by a change in scale.  It is important,
when changing the scale of the image and variance map, that the scale
of the covariance pseudo-matrix also be changed appropriately.  In the
case of binning with an integer scale, this is straightforward.  For
resampling with a non-integer scale, we use bilinear interpolation to
resample the covariance pseudo-matrix.  Unfortunately, this can
introduce a small amount of extra power into the covariance
pseudo-matrix which cannot easily be accounted for --- preserving the
total variance and covariance disturbs the noise model for the
resampled image, while preserving the covariance factor affects the
noise model for convolutions of the resampled image.  In an attempt to
have the best of all worlds, we choose to preserve the total
covariance excluding the variance.

When adding or subtracting (flux) images, the variance maps should be
summed, and the covariance pseudo-matrices combined using a weighted
average, where the weights are the average variances measured from the
variance maps.  This is not perfectly correct for bright sources (nor
can it be, given the limitations of our simple noise model), but this
produces the correct noise for the background, which enables faint
sources to be detected.

\subsection{Example}

\begin{figure}
\plotone{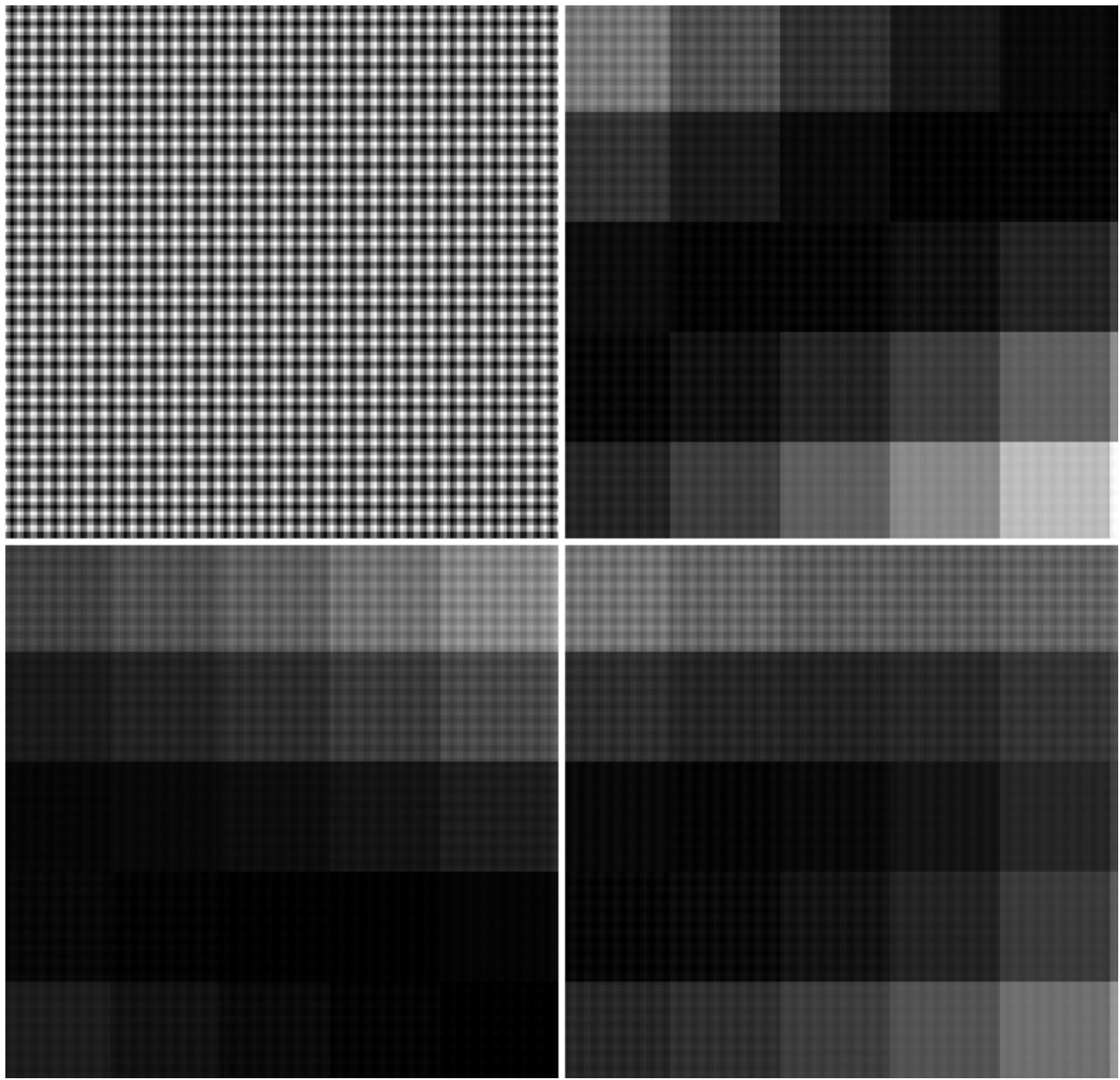}
\caption{Example variance maps from our simulated pipeline.  The
  original (detector frame) variance maps are uninterestingly constant
  and so are not shown.  \textbf{Top left:} Variance map after warping
  (rotation and scale change).  The small-scale structure is due to
  interference between the original and new pixel scales.  \textbf{Top
    right and bottom left:} Variance maps after convolving warped
  images with a spatially variable PSF-matching kernel.  The
  small-scale structure from the warps is still apparent.  The
  moderate-scale squares are due to the application of a constant
  kernel on scales of about $10^2$ pixels, while the large-scale
  structure comes from the spatial variation of the (PSF-matching)
  convolution kernel.  \textbf{Bottom right:} Variance map after
  subtracting the convolved warps.  While the scale and orientation
  for each panel is identical, the color maps are separate.}
\label{fig:noise:variance}
\end{figure}

\begin{figure}
\plotone{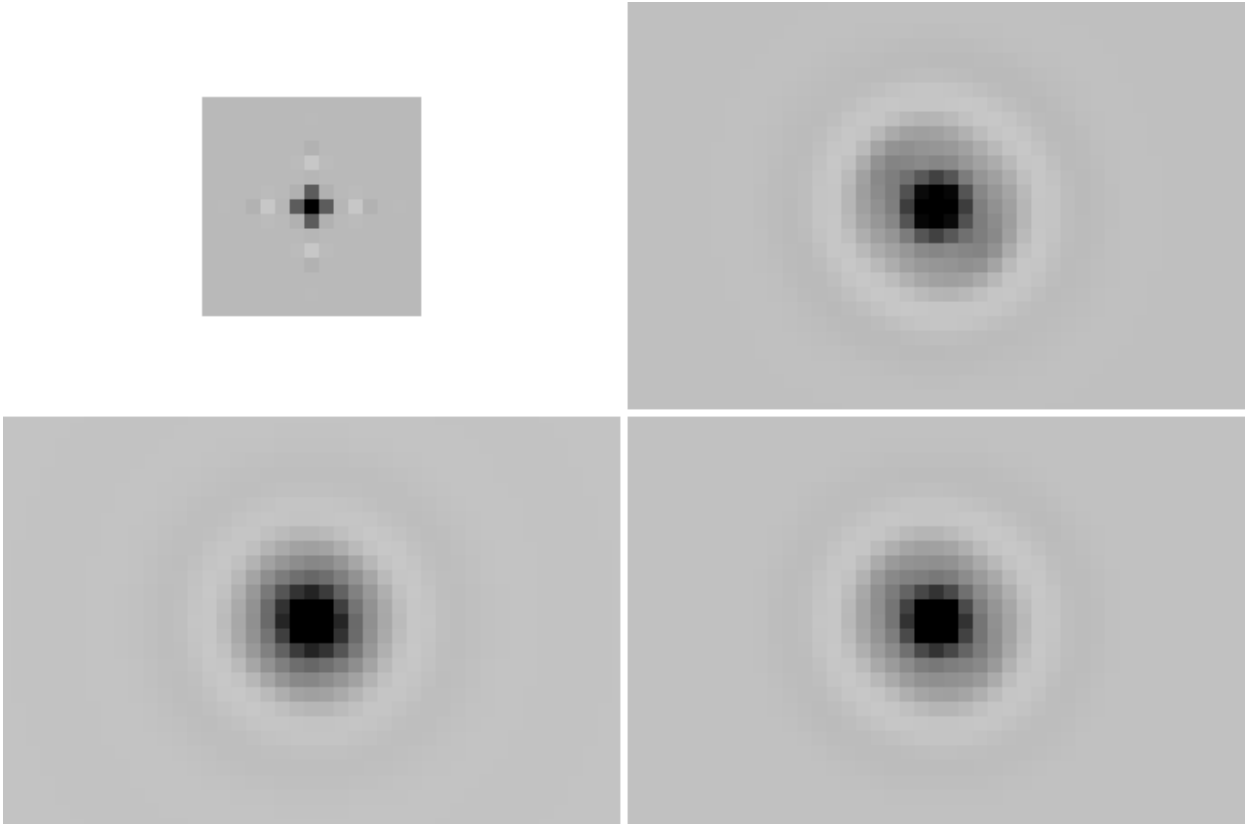}
\caption{Covariance pseudo-matrices from our simulated pipeline.  The
  original (detector frame) covariance pseudo-matrix consists of a
  single pixel of unit value and so is not shown.  The panels are
  arranged in the same manner as Figure~\ref{fig:noise:variance}.
  Warping (in this case, using a {\tt LANCZOS3} interpolation kernel)
  introduces a small amount of covariance, but the PSF-matching
  convolution introduces substantially more.  While the scale and
  orientation for each panel is identical, the color maps are
  separate.}
\label{fig:noise:covariance}
\end{figure}

\begin{figure}
\plottwo{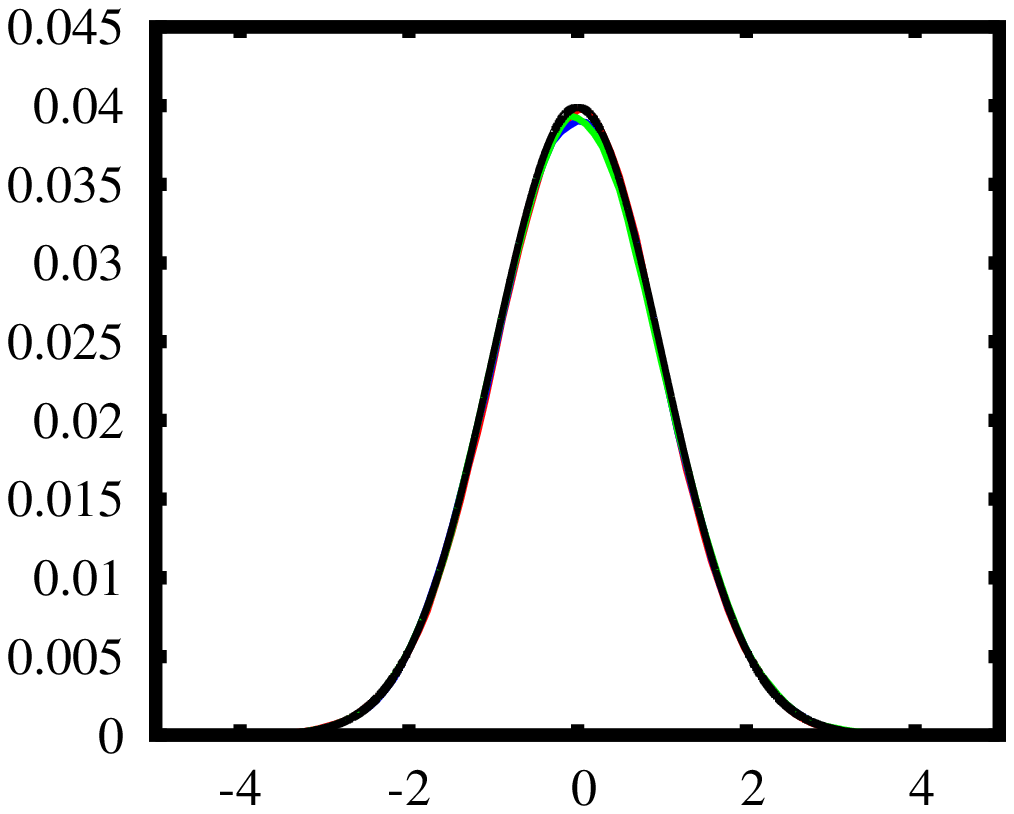}{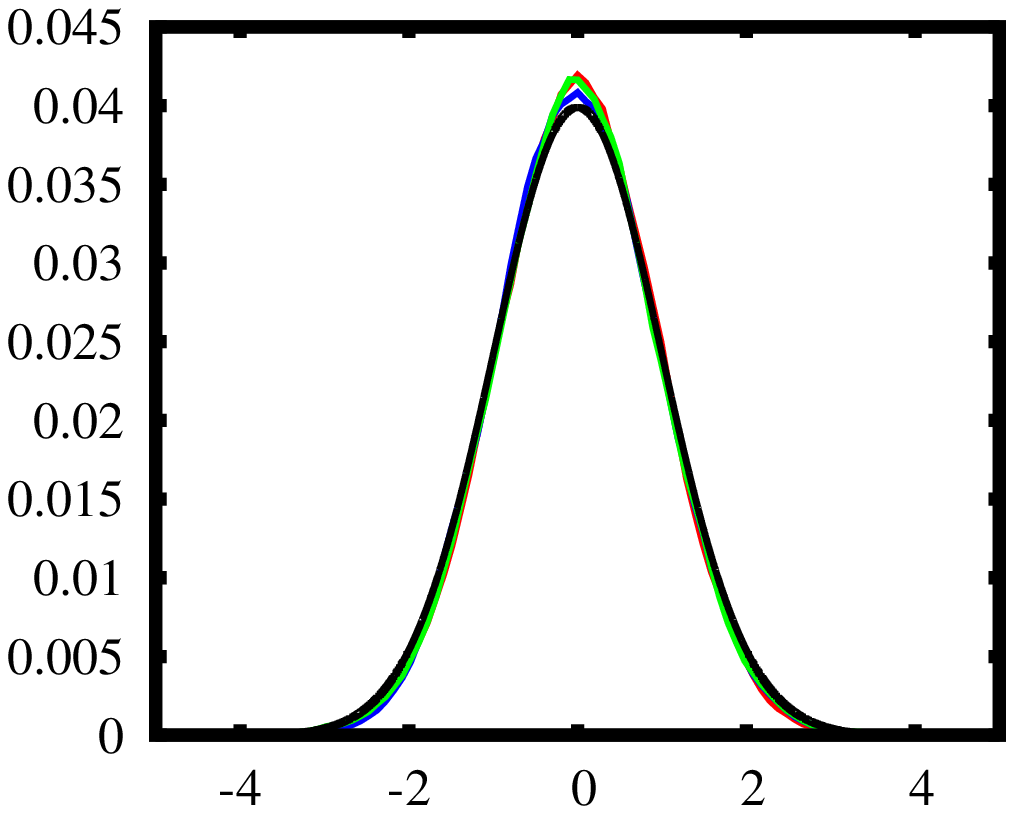}
\plottwo{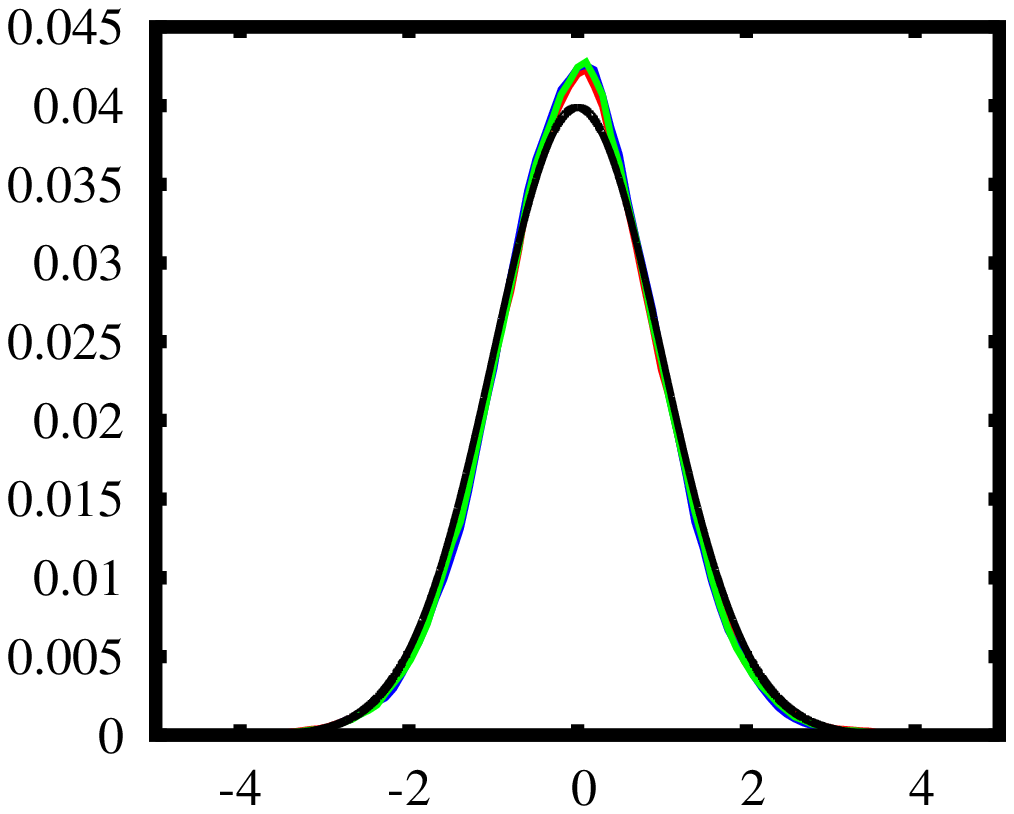}{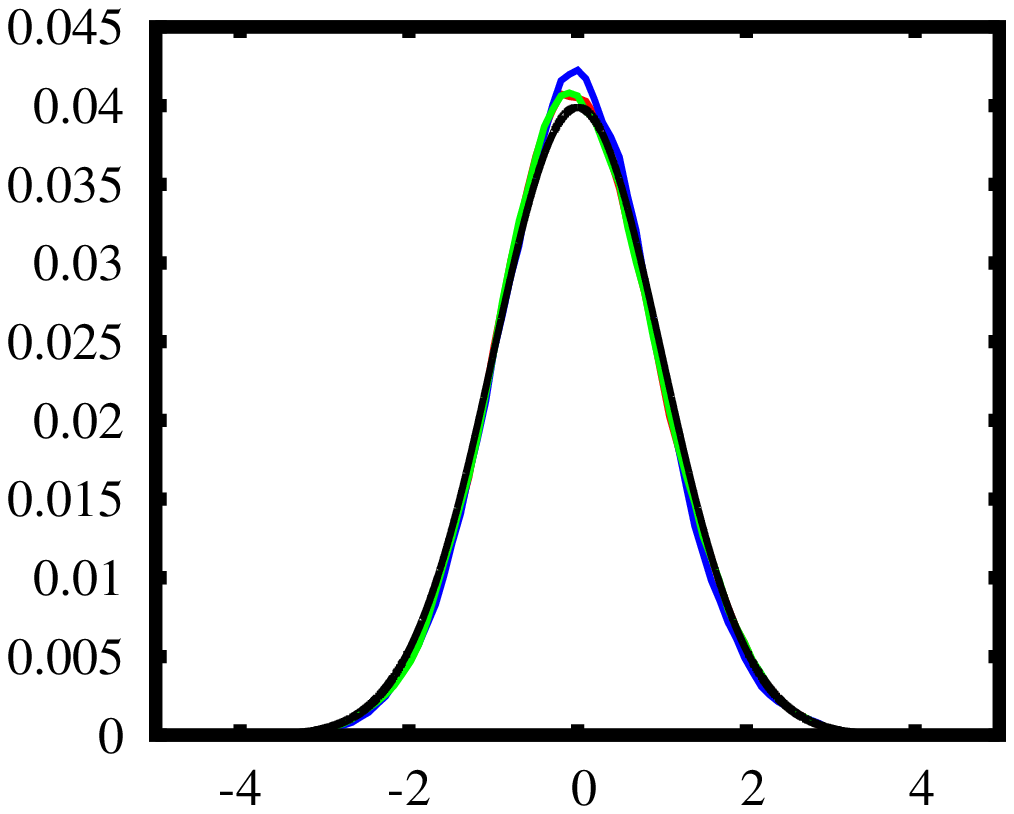}
\caption{Histograms of flux divided by the expected noise after
  convolving with a Gaussian (matched filter for photometry) the
  results of various stages of our simulated pipeline.  In each panel,
  the black line is a Gaussian of unit width and appropriate
  normalisation, while the colored lines (sometimes hidden behind the
  black line) are different realisations from our simulated pipeline.
  \textbf{Top left:} Photometry on the original image.  \textbf{Top
    right:} Photometry on a warped image.  \textbf{Bottom left:}
  Photometry on a subtraction.  \textbf{Bottom right:} Photometry on a
  stack.}
\label{fig:noise:histograms}
\end{figure}

To demonstrate our noise model, we generated fake images composed
entirely of Gaussian noise of different levels.  We then manipulated
these images in a similar manner as for real pipeline operations,
tracking the variance and covariance.  These operations include
warping (rotation and interpolation on a finer pixel scale),
convolution, subtraction of two warped, convolved images, stacking
multiple warped images, subtraction of two convolved stacks, and
subtraction of a convolved warped image from a convolved stacked
image.  Actual PSF-matching kernels were used for the convolutions.
Example variance maps and covariance pseudo-matrices are shown in
Figures~\ref{fig:noise:variance} and \ref{fig:noise:covariance}.  At
each stage, we convolved with a Gaussian, simulating a matched filter
for photometry, and measured the standard deviation of the (flux)
image divided by the square-root of the epected variance, which should
be unity if the noise is being properly tracked.  We find that this
statistic is always within 6\% of unity, with a mean of 0.98 and
r.m.s.\ of 0.03.  Example histograms are shown in
Figure~\ref{fig:noise:histograms}.

\section{Image Stacking}
\label{sec:stacking}

Image stacking is a crucial tool for synoptic surveys, since it allows
long exposures to be traded for multiple visits that can be used to
identify transients, simultaneously satisfying those who desire many
repeated exposures (e.g., for the discovery of transients) and those
who want deep exposures (``wallpaper science'').  Stacking images is
straightforward if the images are all taken under similar conditions,
or with an idealised detector, but neither case is realistic for
ground-based synoptic survey data.  In the presence of masked pixels
(even due to such mundane causes as gaps between devices in a mosaic)
and a variable PSF, a direct combination of images (using any
averaging statistic) will result in an image where the PSF is
spatially variable in a discontinuous fashion, which can introduce
systematic errors to photometry and shape analyses.

Since it would be expensive and unwieldy to calculate the PSF for each
pixel of a stack image separately based on the contributing inputs, we
instead convolve each input to a common PSF (``PSF homogenization'')
using the same PSF-matching technique as for image subtraction,
resulting in a stacked image where the PSF is known everywhere and may
be modeled as varying continuously over the image.  While this
necessarily results in some lost sensitivity due to degrading images
to a wider PSF, this must be balanced against the gain in sensitivity
that comes from controlled systematics.  In any case, it is not
impractical to create stacks both with and without this convolution,
or with a range of PSF widths, allowing the users to choose the most
appropriate for their particular application (e.g., unconvolved stack
for point source detection, convolved stack for analysis of point
sources).

\subsection{Method}

From the known PSF models for the inputs, we determine an ``envelope
PSF'' by realising a circular version of each PSF with a common peak
flux, taking the maximum value pixel by pixel, and fitting the result
with a PSF model.  The envelope PSF may be allowed to vary spatially.
With some more development work, we hope to be able to perform this
envelope calculation directly using the PSF model parameters, and to
reject inputs that would result in a net loss of signal-to-noise ratio
in the stack by their inclusion (due to imposing on all other inputs a
larger PSF envelope than necessary).

To homogenize the PSFs, we create a fake target image using the
envelope PSF and the known positions of (point) sources in the field.
This image is used as a target ($I_2$) for our PSF-matching code
(setting $c_i \equiv 0$) to convolve each image to the common envelope
PSF.  Forcing the envelope PSF to be circular reduces the complexity
of this PSF-matching process, and also when the stack is used as the
input or reference for image subtraction.

Once the inputs have a common PSF, it is a simple matter to identify
outlier pixels (artifacts, transients, etc.) and reject these from
contributing to the stack, without having to worry about the cores of
stars, etc., which may otherwise be a function of the input PSF.  We
flag outlier pixels on a separate mask image, to which we apply a
matched filter (the convolution kernel we applied to the image) to
identify bad pixels on the original image (i.e., before convolution);
this ensures the entire area the bad pixel contributes on the
convolved image is masked, and not just those that were sufficiently
bright to trigger the rejection.  The same list of bad pixels in the
inputs may be used for combining images without convolution.

\subsection{Examples}

As a demonstration of the importance of convolution to match PSFs in
the stack, we created two stacks from observations by PS1 + Giga-Pixel
Camera (GPC1) of the Stellar Transit Survey (STS) field.  We chose
this field because of the high stellar density that will allow a good
comparison of the photometry.  The observations were made under
reasonable conditions (both transparency and seeing) for the purpose
of generating a high-quality stack to be used as the template in image
differencing\footnote{2010 June 3 UT (TJD = 5350), exposure numbers
  0212--0319.}.  The images were processed through the IPP in the
standard manner (dark and flat-field corrections were applied, bad
pixels flagged, photometry and astrometry measured) and the images
were transformed (``warped'') to common reference frames
(``skycells'').  We chose a skycell for which the inputs have a large
amount of overlap.

GPC1 suffers from a condition where pixels that become super-saturated
on an exposure contaminate other pixels within the same column on
subsequent exposures (``burn trails'').  These pixels can easily be
identified and are usually modeled and subtracted fairly well, but the
high stellar density of the STS field affects the quality of the burn
trail subtraction.  We therefore masked all pixels that might contain
a burn trail, but due to the high stellar density and frequent
dithers, this can be a large fraction of the skycell for these images.
This larger than usual masking fraction may amplify any differences
between our convolved stacks and regular (unconvolved) coadditions.

\begin{figure}
\plotone{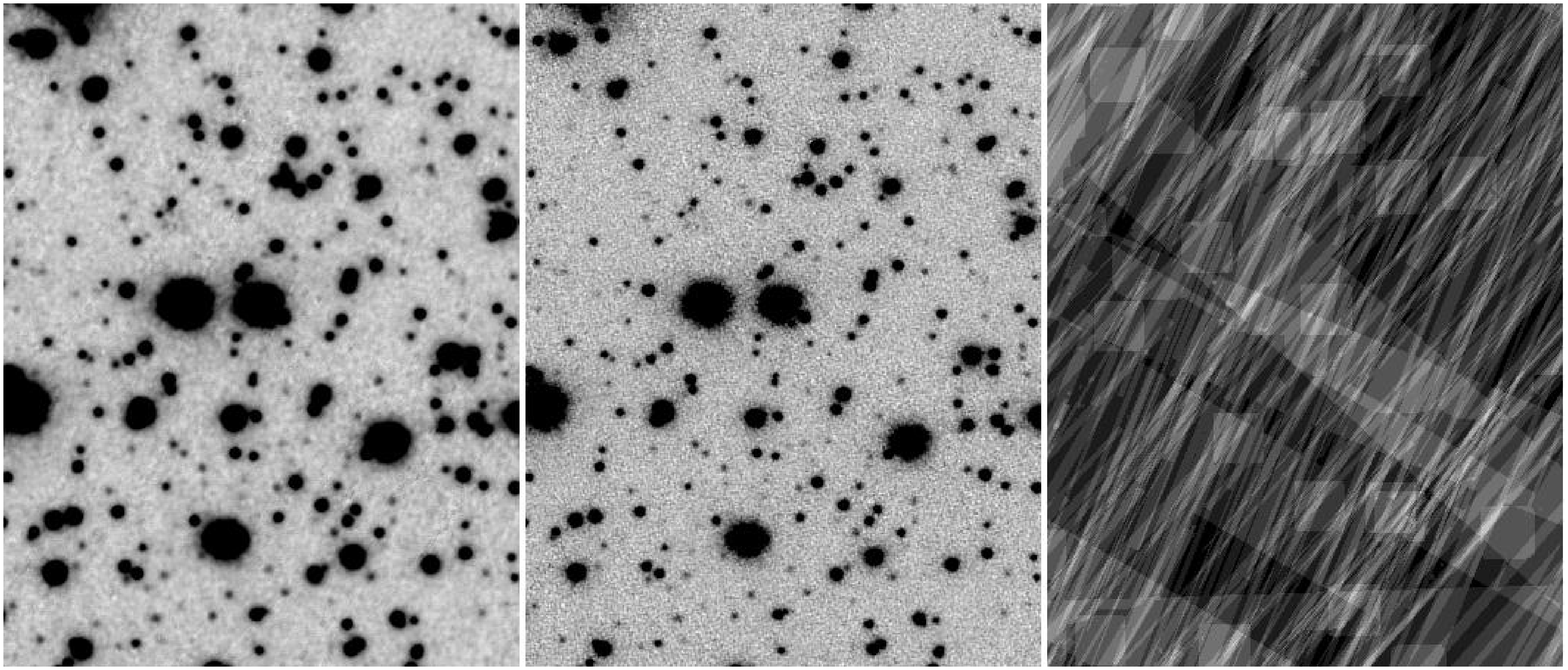}
\caption{Comparison of stacked and unconvolved stacks.  \textbf{Left:}
  Convolved stack.  \textbf{Middle:} Unconvolved
  stack. \textbf{Right:} Exposure map for the convolved stack; note
  the fine structure due to coaddition of masked images.}
\label{fig:stack:image}
\end{figure}

\begin{figure}
\plottwo{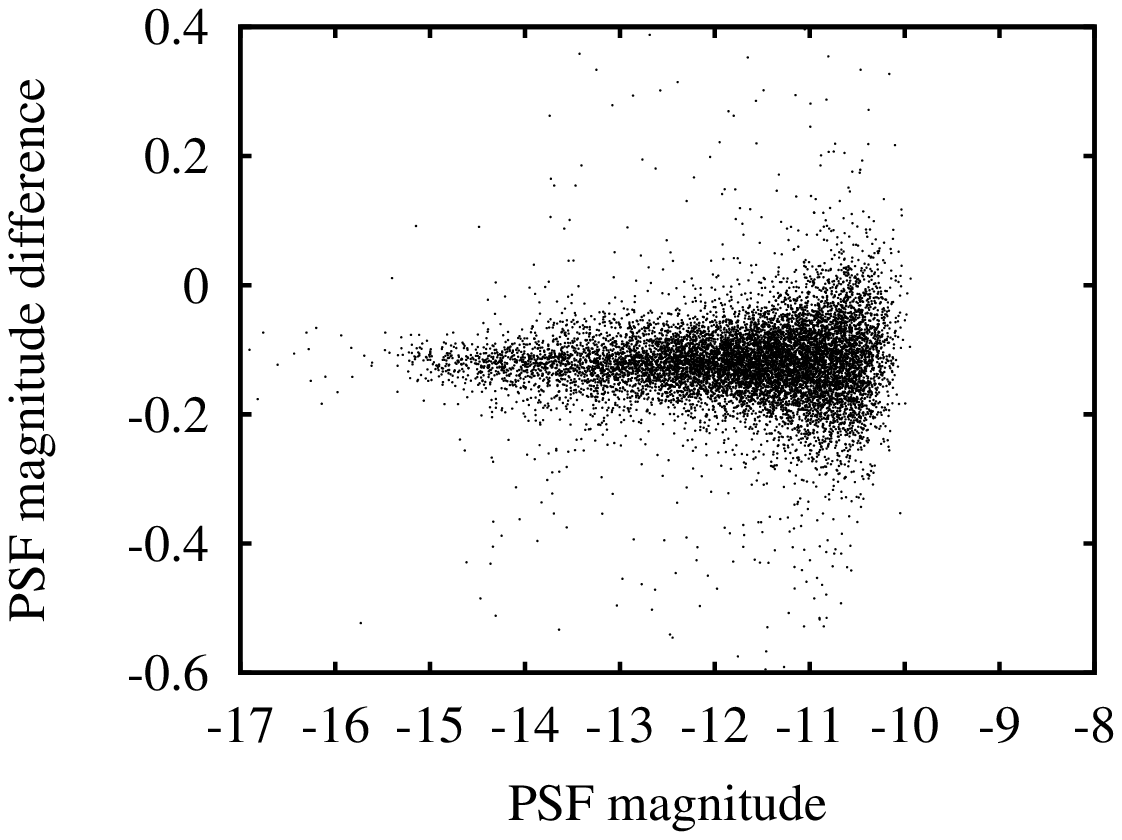}{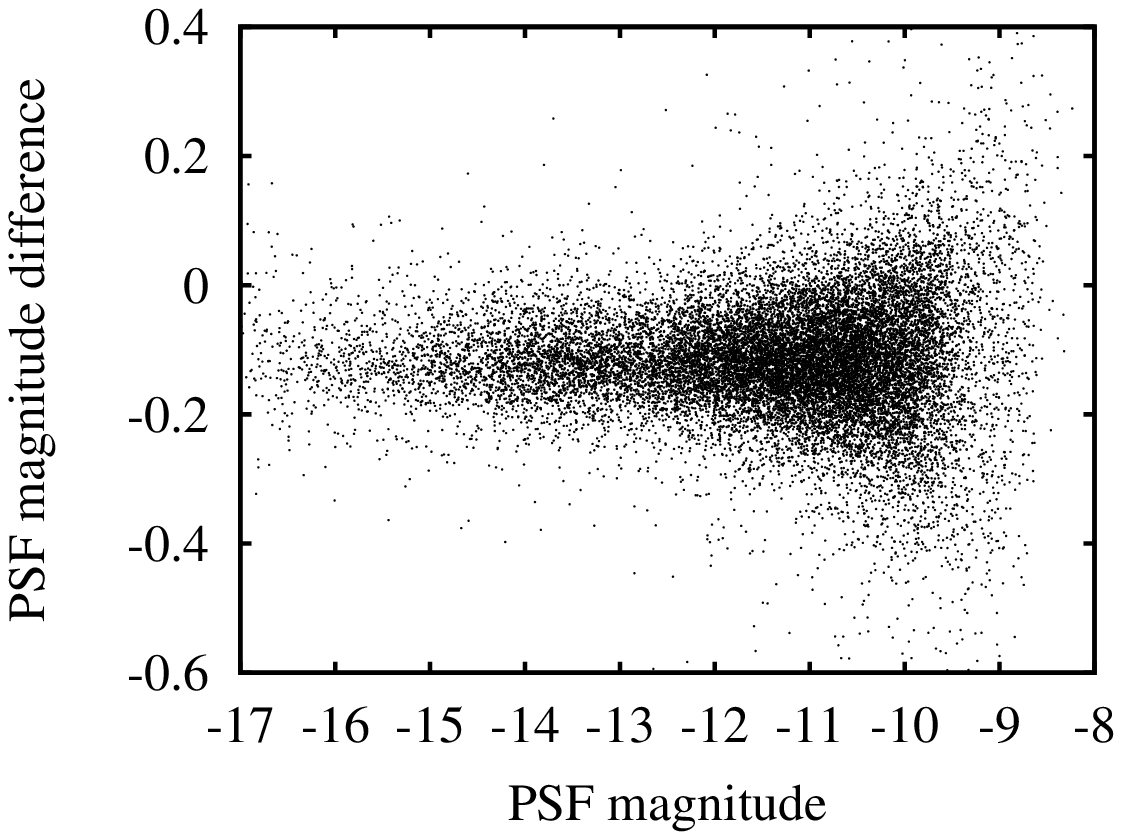}
\plottwo{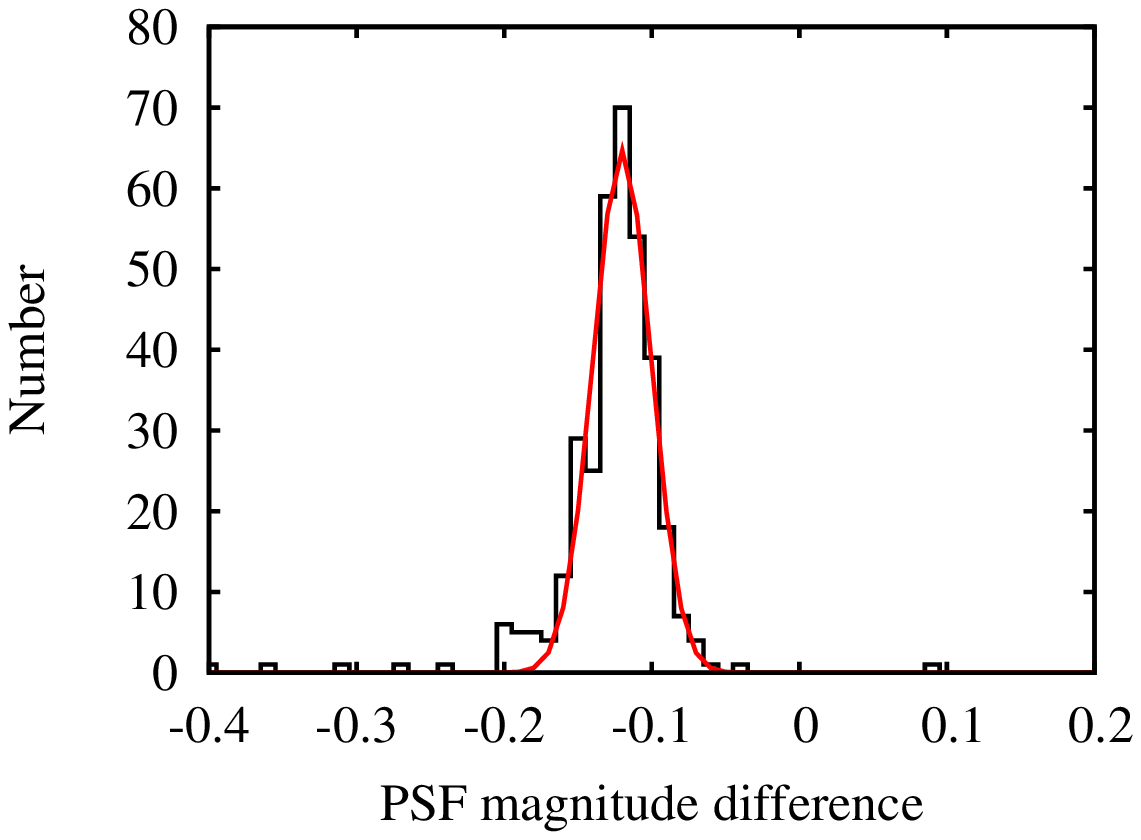}{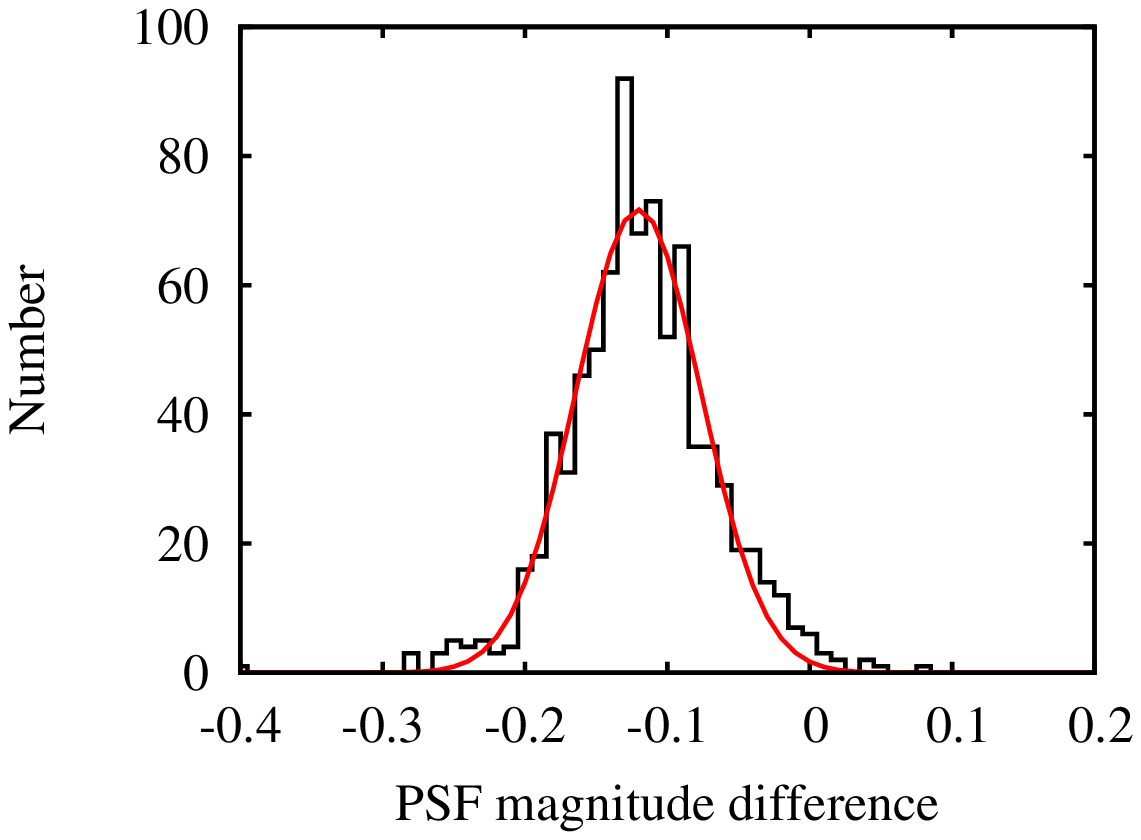}
\caption{Comparison of photometry between convolved and unconvolved
  stacks.  \textbf{Top:} Difference in magnitude as a function of
  (instrumental) magnitude for convolved (left) and unconvolved
  (right) stacks.  \textbf{Bottom:} Histogram of magnitude difference
  for sources with instrumental magnitude between -15 and -14 for
  convolved (left) and unconvolved (right) stacks.  A Gaussian fit has
  been overplotted (red); the widths are 0.020~mag (convolved) and
  0.044~mag (unconvolved).}
\label{fig:stack:phot}
\end{figure}

We chose two disjoint subsets of exposures, each spanning the
available observing time, and stacked them using our stacking program,
\code{ppStack}.  This program generates a convolved stack using the
above method, along with an unconvolved stack using the same pixel
rejection list as for the convolved stack.  In both cases, the pixels
are stacked with weights for each image set to the mean variance in
the convolved image.  A convolved and unconvolved stack are displayed
in Figure~\ref{fig:stack:image}.  We then used our photometry program,
\code{psphot}, to perform PSF photometry on the two convolved and two
unconvolved stacks.  In Figure~\ref{fig:stack:phot} we compare the
magnitude difference as a function of magnitude between the convolved
and unconvolved stacks, along with the distribution of magnitude
differences.  A Gaussian fit to the core of each distribution yields a
width of 0.020~mag for the convolved stack and 0.044~mag for the
unconvolved stack, demonstrating the need for PSF homogenisation for
accurate stack photometry.  Of course, the convolved stacks are not as
deep as the unconvolved stacks, but one could easily imagine a hybrid
photometry scheme where bright objects are measured from the convolved
stack while faint objects are measured from the unconvolved stack, or
detected on the unconvolved stack but measured on the convolved stack.

\section{Putting it all together}
\label{sec:all}

We are routinely running our image stacking and subtraction codes on
images from the Pan-STARRS~1 telescope, from which SN discoveries are
being made \citep[e.g., see][]{2010ApJ...717L..52B}.

As a demonstration of the above techniques, we re-create the discovery
of SN~2009kf \citep{2010ApJ...717L..52B}, showing each step along the
way.  Our input data consists of 8 $r$-band exposures of PS1
Medium-Deep field 08 (MD08) from before the discovery and 4 $r$-band
exposures from the discovery epoch\footnote {2009 May 22 UT (TJD =
  4973), exposure numbers 0126--0133; and 2009 June 11 UT (TJD =
  4993), exposure numbers 0088--0091.}.  The images were processed
through the IPP in the standard manner and warped to skycells.

\begin{figure}
\plotone{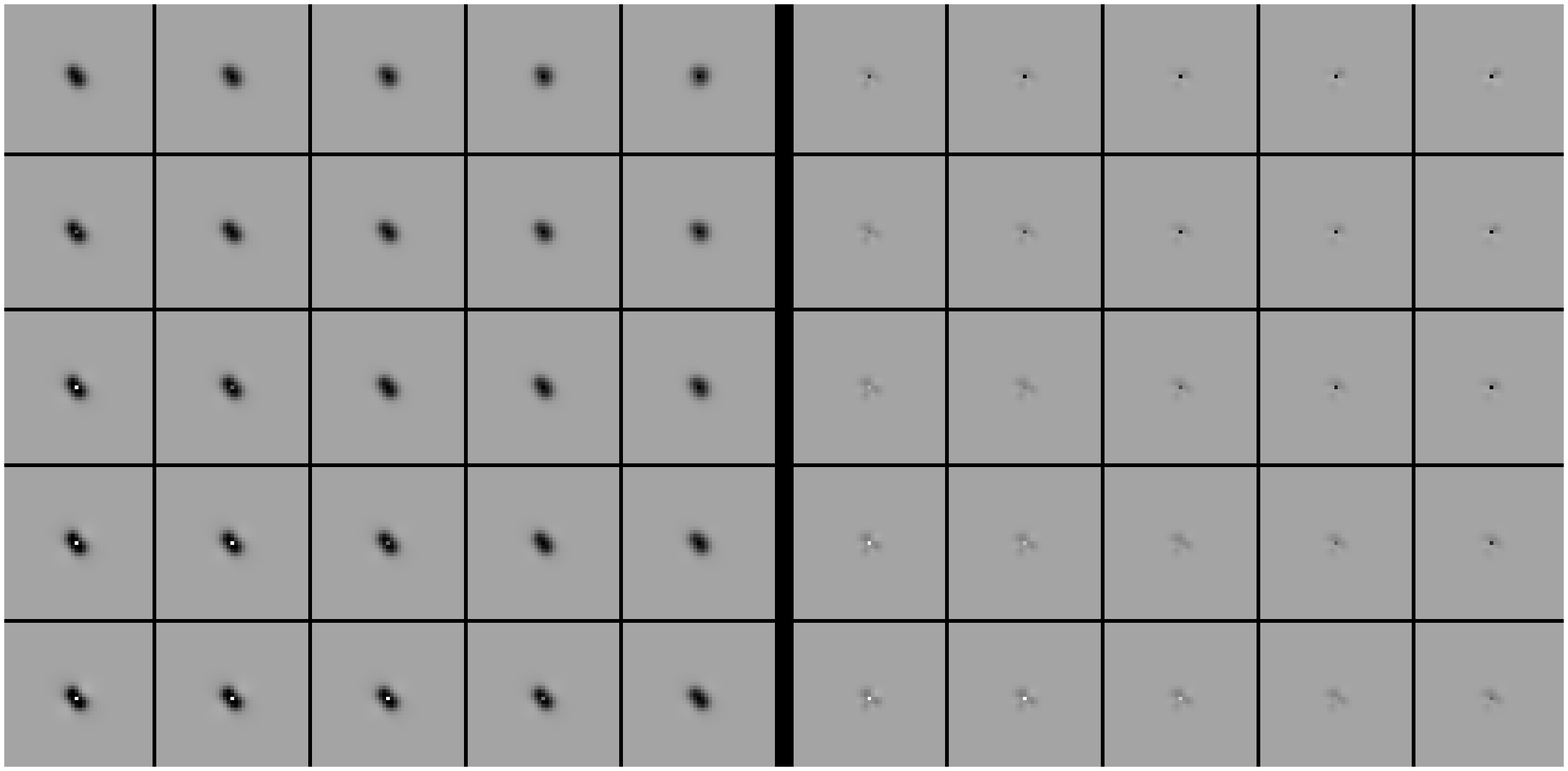}
\caption{Kernels for PSF-matching an individual exposure to the
  reference stack.  \textbf{Left:} Kernel applied to the individual
  exposure ($K_1$).  \textbf{Right:} Kernel applied to the reference
  stack ($K_2$).  For each, the kernel is realised at a grid of points
  over the image, so that the spatial variation can be visualised.}
\label{fig:all:kernel}
\end{figure}

We stacked the exposures from before the discovery to use as our
template.  We subtracted this reference from each of the individual
warps from the discovery epoch.  Example PSF-matching kernels are
shown in Figure~\ref{fig:all:kernel}.  It is apparent that the kernels
are strongly spatially variable.  In this case, the second kernel
($K_2$) does not play a major part in the PSF-matching (because the
reference stack PSF has been circularised as part of the PSF
homogenisation process), but it does contain a small amount of power
that helps produce a good subtraction.

\begin{figure}
\plotone{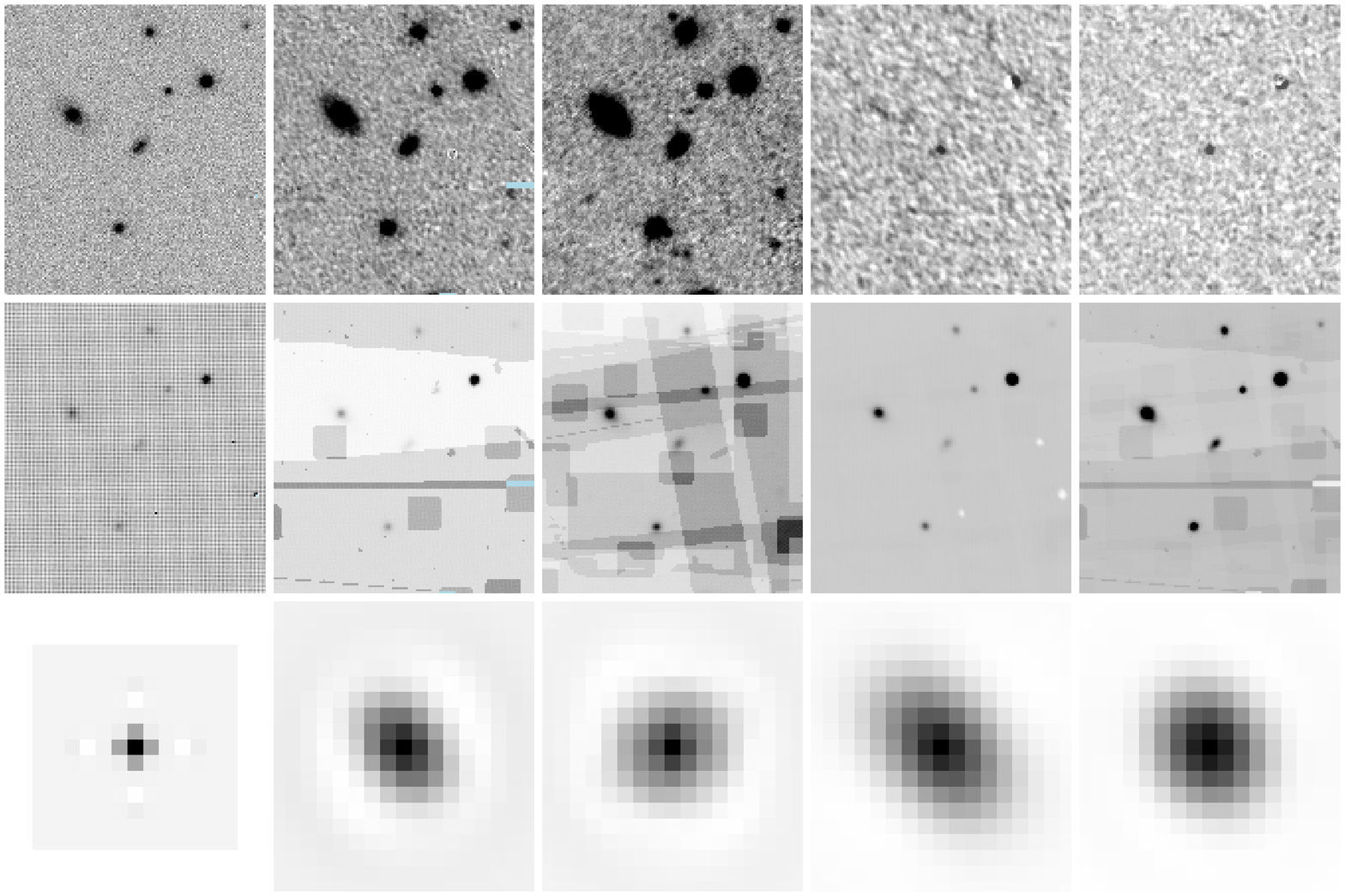}
\caption{Recreation of the discovery of SN~2009kf
  \citep{2010ApJ...717L..52B}, demonstrating the techniques from this
  paper.  Across a row, the images are an individual exposure, stack
  of the discovery epoch exposures, the reference stack, subtraction
  of the reference from the individual exposure, and subtraction of
  the reference from the discovery epoch stack.  The rows are the flux
  (top), the variance map (middle) and the covariance pseudo-matrix
  (bottom).  The orientation of the flux and variance maps is
  identical, but the color maps are not.  The covariance maps are
  displayed at a common scale, but with different color maps.  The SN
  is well-detected in both the individual and stack subtractions.}
\label{fig:all:images}
\end{figure}

We also stacked the discovery epoch warps and subtracted from this
stack the template.  The SN is well-detected in all cases
(Figure~\ref{fig:all:images}).  The variance maps for the stacks are
complicated due to chip gaps and the rejection of artifacts, and this
structure is propagated to the subtracted images.  Following
subtraction, the covariance is extended by as much as FWHM $\sim$
5~pixels, and in some cases, elongated.  Because of these effects,
photometry of variable sources such as SNe and asteroids that does not
take into account the variance structure and covariance is likely to
mis-estimate the photometric errors.

The algorithms presented here have been implemented within the PS1
IPP, which is freely available from our Subversion
repository\footnote{\url{http://svn.pan-starrs.ifa.hawaii.edu/repo/ipp/}}.
The IPP has been running routinely on data collected by PS1 since
February 2010, resulting in the discovery of hundreds of SNe and
thousands of asteroids as PS1 surveys the heavens.

\acknowledgments

We thank Robert Lupton and Andy Becker for useful discussions about
image subtraction; and Armin Rest, Michael Wood-Vasey, Mark Huber,
Maria-Theresa Botticella and Stefano Valenti for testing the
subtraction code, and Nigel Metcalfe and Peter Draper for testing the
stack code.  We also thank our fellow IPP team members Bill Sweeney,
Josh Hoblitt, Heather Flewelling and Chris Waters for their parts in
producing the IPP.  PAP thanks Brian Schmidt for first teaching him
the technique of image subtraction.  PAP uses and recommends the
SAOImage DS9 (developed by Smithsonian Astrophysical Observatory) and
TOPCAT (by Mark Taylor) software tools.  The PS1 Surveys have been
made possible through the combinations of the Institute for Astronomy
at the University of Hawaii, The Pan-STARRS Project Office, the
Max-Planck Society and its participating institutes, the Max Planck
Institute for Astronomy, Heidelberg, and the Max Planck Institute for
Extraterrestial Physics, Garching, The Johns Hopkins University, the
University of Durham, the University of Edinburgh, the Queen's
University of Belfast, the Harvard-Smithsonian Center for
Astrophysics, the Las Cumbres Observatory Global Network, and the
National Central University of Taiwan.

\facility{Pan-STARRS 1}
\clearpage
\bibliographystyle{apj}
%\bibliography{apj-jour,references}

\begin{thebibliography}{16}
\expandafter\ifx\csname natexlab\endcsname\relax\def\natexlab#1{#1}\fi

\bibitem[{{Akerlof} {et~al.}(2000){Akerlof}, {Balsano}, {Barthelmy}, {Bloch},
  {Butterworth}, {Casperson}, {Cline}, {Fletcher}, {Gisler}, {Hills}, {Kehoe},
  {Lee}, {Marshall}, {McKay}, {Pawl}, {Priedhorsky}, {Seldomridge},
  {Szymanski}, \& {Wren}}]{2000ApJ...542..251A}
{Akerlof}, C., {Balsano}, R., {Barthelmy}, S., {Bloch}, J., {Butterworth}, P.,
  {Casperson}, D., {Cline}, T., {Fletcher}, S., {Gisler}, G., {Hills}, J.,
  {Kehoe}, R., {Lee}, B., {Marshall}, S., {McKay}, T., {Pawl}, A.,
  {Priedhorsky}, W., {Seldomridge}, N., {Szymanski}, J., \& {Wren}, J. 2000,
  \apj, 542, 251

\bibitem[{{Alard}(2000)}]{2000AAS..144..363A}
{Alard}, C. 2000, \aaps, 144, 363

\bibitem[{{Alard} \& {Lupton}(1998)}]{1998ApJ...503..325A}
{Alard}, C. \& {Lupton}, R.~H. 1998, \apj, 503, 325

\bibitem[{{Alcock} {et~al.}(1993){Alcock}, {Allsman}, {Axelrod}, {Bennett},
  {Cook}, {Park}, {Marshall}, {Stubbs}, {Griest}, {Perlmutter}, {Sutherland},
  {Freeman}, {Peterson}, {Quinn}, \& {Rodgers}}]{1993ASPC...43..291A}
{Alcock}, C., {Allsman}, R.~A., {Axelrod}, T.~S., {Bennett}, D.~P., {Cook},
  K.~H., {Park}, H.~S., {Marshall}, S.~L., {Stubbs}, C.~W., {Griest}, K.,
  {Perlmutter}, S., {Sutherland}, W., {Freeman}, K.~C., {Peterson}, B.~A.,
  {Quinn}, P.~J., \& {Rodgers}, A.~W. 1993, in Astronomical Society of the
  Pacific Conference Series, Vol.~43, Sky Surveys. Protostars to Protogalaxies,
  ed. B.~T. {Soifer}, 291

\bibitem[{{Bertin} {et~al.}(2002){Bertin}, {Mellier}, {Radovich}, {Missonnier},
  {Didelon}, \& {Morin}}]{2002ASPC..281..228B}
{Bertin}, E., {Mellier}, Y., {Radovich}, M., {Missonnier}, G., {Didelon}, P.,
  \& {Morin}, B. 2002, in Astronomical Society of the Pacific Conference
  Series, Vol. 281, Astronomical Data Analysis Software and Systems XI, ed.
  {D.~A.~Bohlender, D.~Durand, \& T.~H.~Handley}, 228--+

\bibitem[{{Botticella} {et~al.}(2010){Botticella}, {Trundle}, {Pastorello},
  {Rodney}, {Rest}, {Gezari}, {Smartt}, {Narayan}, {Huber}, {Tonry}, {Young},
  {Smith}, {Bresolin}, {Valenti}, {Kotak}, {Mattila}, {Kankare}, {Wood-Vasey},
  {Riess}, {Neill}, {Forster}, {Martin}, {Stubbs}, {Burgett}, {Chambers},
  {Dombeck}, {Flewelling}, {Grav}, {Heasley}, {Hodapp}, {Kaiser}, {Kudritzki},
  {Luppino}, {Lupton}, {Magnier}, {Monet}, {Morgan}, {Onaka}, {Price},
  {Rhoads}, {Siegmund}, {Sweeney}, {Wainscoat}, {Waters}, {Waterson}, \&
  {Wynn-Williams}}]{2010ApJ...717L..52B}
{Botticella}, M.~T., {Trundle}, C., {Pastorello}, A., {Rodney}, S., {Rest}, A.,
  {Gezari}, S., {Smartt}, S.~J., {Narayan}, G., {Huber}, M.~E., {Tonry}, J.~L.,
  {Young}, D., {Smith}, K., {Bresolin}, F., {Valenti}, S., {Kotak}, R.,
  {Mattila}, S., {Kankare}, E., {Wood-Vasey}, W.~M., {Riess}, A., {Neill},
  J.~D., {Forster}, K., {Martin}, D.~C., {Stubbs}, C.~W., {Burgett}, W.~S.,
  {Chambers}, K.~C., {Dombeck}, T., {Flewelling}, H., {Grav}, T., {Heasley},
  J.~N., {Hodapp}, K.~W., {Kaiser}, N., {Kudritzki}, R., {Luppino}, G.,
  {Lupton}, R.~H., {Magnier}, E.~A., {Monet}, D.~G., {Morgan}, J.~S., {Onaka},
  P.~M., {Price}, P.~A., {Rhoads}, P.~H., {Siegmund}, W.~A., {Sweeney}, W.~E.,
  {Wainscoat}, R.~J., {Waters}, C., {Waterson}, M.~F., \& {Wynn-Williams},
  C.~G. 2010, \apjl, 717, L52

\bibitem[{{Bowell} {et~al.}(1995){Bowell}, {Koehn}, {Howell}, {Hoffman}, \&
  {Muinonen}}]{1995DPS....27.0110B}
{Bowell}, E., {Koehn}, B.~W., {Howell}, S.~B., {Hoffman}, M., \& {Muinonen}, K.
  1995, in Bulletin of the American Astronomical Society, Vol.~27, AAS/Division
  for Planetary Sciences Meeting Abstracts \#27, 1057

\bibitem[{{Bramich}(2008)}]{2008MNRAS.386L..77B}
{Bramich}, D.~M. 2008, \mnras, 386, L77

\bibitem[{{Chambers} {et~al.}(2017){Chambers}, {Magnier}, {Metcalfe}, \&
  et~al.}]{chambers2017}
{Chambers}, K.~C., {Magnier}, E.~A., {Metcalfe}, N., \& et~al. 2017, ArXiv
  e-prints

\bibitem[{{Larson} {et~al.}(2003){Larson}, {Beshore}, {Hill}, {Christensen},
  {McLean}, {Kolar}, {McNaught}, \& {Garradd}}]{2003DPS....35.3604L}
{Larson}, S., {Beshore}, E., {Hill}, R., {Christensen}, E., {McLean}, D.,
  {Kolar}, S., {McNaught}, R., \& {Garradd}, G. 2003, in Bulletin of the
  American Astronomical Society, Vol.~35, AAS/Division for Planetary Sciences
  Meeting Abstracts \#35, 982

\bibitem[{{Law} {et~al.}(2009){Law}, {Kulkarni}, {Dekany}, {Ofek}, {Quimby},
  {Nugent}, {Surace}, {Grillmair}, {Bloom}, {Kasliwal}, {Bildsten}, {Brown},
  {Cenko}, {Ciardi}, {Croner}, {Djorgovski}, {van Eyken}, {Filippenko}, {Fox},
  {Gal-Yam}, {Hale}, {Hamam}, {Helou}, {Henning}, {Howell}, {Jacobsen},
  {Laher}, {Mattingly}, {McKenna}, {Pickles}, {Poznanski}, {Rahmer}, {Rau},
  {Rosing}, {Shara}, {Smith}, {Starr}, {Sullivan}, {Velur}, {Walters}, \&
  {Zolkower}}]{2009PASP..121.1395L}
{Law}, N.~M., {Kulkarni}, S.~R., {Dekany}, R.~G., {Ofek}, E.~O., {Quimby},
  R.~M., {Nugent}, P.~E., {Surace}, J., {Grillmair}, C.~C., {Bloom}, J.~S.,
  {Kasliwal}, M.~M., {Bildsten}, L., {Brown}, T., {Cenko}, S.~B., {Ciardi}, D.,
  {Croner}, E., {Djorgovski}, S.~G., {van Eyken}, J., {Filippenko}, A.~V.,
  {Fox}, D.~B., {Gal-Yam}, A., {Hale}, D., {Hamam}, N., {Helou}, G., {Henning},
  J., {Howell}, D.~A., {Jacobsen}, J., {Laher}, R., {Mattingly}, S., {McKenna},
  D., {Pickles}, A., {Poznanski}, D., {Rahmer}, G., {Rau}, A., {Rosing}, W.,
  {Shara}, M., {Smith}, R., {Starr}, D., {Sullivan}, M., {Velur}, V.,
  {Walters}, R., \& {Zolkower}, J. 2009, \pasp, 121, 1395

\bibitem[{{Shappee} {et~al.}(2014){Shappee}, {Prieto}, {Grupe}, {Kochanek},
  {Stanek}, {De Rosa}, {Mathur}, {Zu}, {Peterson}, {Pogge}, {Komossa}, {Im},
  {Jencson}, {Holoien}, {Basu}, {Beacom}, {Szczygie{\l}}, {Brimacombe},
  {Adams}, {Campillay}, {Choi}, {Contreras}, {Dietrich}, {Dubberley},
  {Elphick}, {Foale}, {Giustini}, {Gonzalez}, {Hawkins}, {Howell}, {Hsiao},
  {Koss}, {Leighly}, {Morrell}, {Mudd}, {Mullins}, {Nugent}, {Parrent},
  {Phillips}, {Pojmanski}, {Rosing}, {Ross}, {Sand}, {Terndrup}, {Valenti},
  {Walker}, \& {Yoon}}]{2014ApJ...788...48S}
{Shappee}, B.~J., {Prieto}, J.~L., {Grupe}, D., {Kochanek}, C.~S., {Stanek},
  K.~Z., {De Rosa}, G., {Mathur}, S., {Zu}, Y., {Peterson}, B.~M., {Pogge},
  R.~W., {Komossa}, S., {Im}, M., {Jencson}, J., {Holoien}, T.~W.-S., {Basu},
  U., {Beacom}, J.~F., {Szczygie{\l}}, D.~M., {Brimacombe}, J., {Adams}, S.,
  {Campillay}, A., {Choi}, C., {Contreras}, C., {Dietrich}, M., {Dubberley},
  M., {Elphick}, M., {Foale}, S., {Giustini}, M., {Gonzalez}, C., {Hawkins},
  E., {Howell}, D.~A., {Hsiao}, E.~Y., {Koss}, M., {Leighly}, K.~M., {Morrell},
  N., {Mudd}, D., {Mullins}, D., {Nugent}, J.~M., {Parrent}, J., {Phillips},
  M.~M., {Pojmanski}, G., {Rosing}, W., {Ross}, R., {Sand}, D., {Terndrup},
  D.~M., {Valenti}, S., {Walker}, Z., \& {Yoon}, Y. 2014, \apj, 788, 48

\bibitem[{{Stokes} {et~al.}(2000){Stokes}, {Evans}, {Viggh}, {Shelly}, \&
  {Pearce}}]{2000Icar..148...21S}
{Stokes}, G.~H., {Evans}, J.~B., {Viggh}, H.~E.~M., {Shelly}, F.~C., \&
  {Pearce}, E.~C. 2000, \icarus, 148, 21

\bibitem[{{Tonry} {et~al.}(2018){Tonry}, {Denneau}, {Heinze}, {Stalder},
  {Smith}, {Smartt}, {Stubbs}, {Weiland}, \& {Rest}}]{2018PASP..130f4505T}
{Tonry}, J.~L., {Denneau}, L., {Heinze}, A.~N., {Stalder}, B., {Smith}, K.~W.,
  {Smartt}, S.~J., {Stubbs}, C.~W., {Weiland}, H.~J., \& {Rest}, A. 2018,
  \pasp, 130, 064505

\bibitem[{{Udalski} {et~al.}(1992){Udalski}, {Szymanski}, {Kaluzny}, {Kubiak},
  \& {Mateo}}]{1992AcA....42..253U}
{Udalski}, A., {Szymanski}, M., {Kaluzny}, J., {Kubiak}, M., \& {Mateo}, M.
  1992, \actaa, 42, 253

\bibitem[{{Yuan} \& {Akerlof}(2008)}]{2008ApJ...677..808Y}
{Yuan}, F. \& {Akerlof}, C.~W. 2008, \apj, 677, 808

\end{thebibliography}

\end{document}